\documentclass[11pt,a4paper]{article} 
 
        \usepackage[utf8]{inputenc}
        \usepackage{fontenc}
	\usepackage[italian,english]{babel}
	\usepackage{amsmath}
	\usepackage{geometry}
	\usepackage{amsthm}
	\usepackage{mathrsfs}
	\usepackage{bbold}
	\usepackage{mathtools}
	\usepackage{amsfonts}
	\usepackage{graphicx}
	\usepackage{setspace}
	\usepackage{physics}
	 \usepackage{comment}
	  \usepackage{layout}
	\usepackage{xcolor}
     \usepackage[big]{layaureo}
	 \usepackage{dsfont}	
	 \usepackage{slashed}
 	 \usepackage{amssymb}
	 \usepackage{tensor}
	\usepackage{fancyhdr} 
 	\usepackage{tikz-cd}
 	\usepackage{braket}
 	\usepackage{tikz}
 	\usepackage[compat=1.1.0]{tikz-feynman}
 	\usepackage{subfigure}
	\usepackage{fancyhdr} 
	\usepackage{booktabs}
	\usepackage{braket}
	
\renewcommand{\arraystretch}{1.5}
	        
\usepackage{jheppub} 

\usepackage{natbib}
\title{The topological line of ABJ(M) theory}
\author[a]{Nicola Gorini,} 
\author[b]{Luca Griguolo,} 
\author[b]{Luigi Guerrini,} 
\author[a]{Silvia Penati,}
\author[c]{Domenico Seminara,}
\author[b]{Paolo Soresina}

\affiliation[a]{ Dipartimento di Fisica, Universit\`a degli studi di Milano--Bicocca, and INFN, Sezione di Milano--Bicocca, Piazza della Scienza 3, I-20126 Milano, Italy }
\affiliation[b]{Dipartimento SMFI, Universit\`a di Parma and INFN Gruppo Collegato di Parma, Viale G.P. Usberti 7/A, 43100 Parma, Italy}
\affiliation[c]{Dipartimento di Fisica, Universit\`a di Firenze and INFN Sezione di Firenze, via G. Sansone 1, 50019 Sesto Fiorentino, Italy\\}

\emailAdd{n.gorini1@campus.unimib.it}  
\emailAdd{luca.griguolo@pr.infn.it} 
\emailAdd{luigi.guerrini@unipr.it} 
\emailAdd{silvia.penati@mib.infn.it} 
\emailAdd{seminara@fi.infn.it}
\emailAdd{paolo.soresina@unipr.it}

\abstract{We construct the one-dimensional topological sector of $\mathcal N=6$ ABJ(M) theory and study its relation with the mass-deformed partition function on $S^3$. Supersymmetric localization provides an exact representation of this partition function as a matrix integral, which interpolates between weak and strong coupling regimes. It has been proposed that correlation functions of dimension-one topological operators should be computed through suitable derivatives with respect to the masses, but a precise proof is still lacking. We present non-trivial evidence for this relation by computing the two-point function at two-loop, successfully matching the matrix model expansion at weak coupling and finite ranks. As a by-product we obtain the two-loop explicit expression for the central charge $c_T$ of ABJ(M) theory.  Three- and four-point functions up to one-loop confirm the relation as well. Our result points towards the possibility to localize the one-dimensional topological sector of ABJ(M) and may also be useful in the bootstrap program for 3d SCFTs.}

	\newcommand{\beq}{\begin{equation}}
	\newcommand{\bea}{\begin{eqnarray}}
	\newcommand{\eea}{\end{eqnarray}}
	\newcommand{\eeq}{\end{equation}}
	\newcommand{\non}{\nonumber}

	\newcommand{\e}{\mathrm{e}}

	\newcommand{\btheta}{\bar{\theta}}

	\newcommand{\bQ}{\bar{Q}}

	\newcommand{\bY}{\bar{Y}}
	\newcommand{\bZ}{\bar{Z}}

	\newcommand{\de}{\partial}
	\newcommand{\Op}{\mathcal{O}}
	\renewcommand{\a}{\alpha}
	\renewcommand{\b}{\beta}
	\newcommand{\g}{\gamma}
	\renewcommand{\e}{\epsilon}
	\newcommand{\pa}{\partial}

	\makeatletter
	\newcommand{\aextp}{\@ifnextchar^\@aextp{\@aextp^{\,}}}
	\def\@aextp^#1{\mathop{\bigwedge\nolimits^{\!#1}}}
	\makeatother
	
	\makeatletter
	\newcommand{\extp}{\@ifnextchar_\@extp{\@extp_{\,}}}
	\def\@extp_#1{\mathop{\aextp\nolimits_{\!#1}}}
	\makeatother

	\theoremstyle{definition}

\begin{document}
\maketitle

\section{Introduction}

${\cal N} = 6$ superconformal field theories (SCFTs) play a pivotal role in studying the superconformal window of quantum field theory in three dimensions, thanks to crucial properties that are worth to be emphasized. 
First of all, they provide an explicit realization of the AdS/CFT correspondence \cite{Maldacena:1997re,Gubser:1998bc,Witten:1998qj}, since they possess string or M-theory duals, or  weakly coupled higher-spin fields in $\text{AdS}_4$ \cite{Chang:2012kt}. The best known example is the class of ABJ(M) quiver theories with gauge groups $U(N_1)_k\times U(N_2)_{-k}$, being $k$ the Chern-Simons level, which are dual to M-theory on AdS$_4 \times S^7/{\mathbb Z}_k$  or type IIA string theory on AdS$_4 \times {\rm CP}^3$, depending on the particular range of the coupling constants \cite{Aharony:2008ug,Aharony:2008gk}. The other crucial property is that this amount of supersymmetry, while being not too restrictive and thus allowing for a large family of such SCFTs, is somehow sufficient for attacking the non-perturbative regime of the theory via the exact evaluation of various protected quantities.

From a general point of view, progress on the non-perturbative dynamics of SCFTs has been favoured by two powerful tools, the conformal bootstrap and supersymmetric localization. The conformal bootstrap method has revealed to be a very successful tool in obtaining exact results in CFTs, including the famous 3D Ising model \cite{ElShowk:2012ht}. The inclusion of supersymmetry and the combination of bootstrap techniques with supersymmetric localization \cite{Pestun:2016zxk,Kapustin:2009kz} has made analytic computations in SCFTs possible,  as shown for example in \cite{Chester:2014fya,Chester:2014mea,Chang:2017cdx,Chang:2017xmr,Baggio:2017mas,Agmon:2017xes,Agmon:2019imm}. In these advances, a prominent role has been played by topological sectors, consisting of a completely solvable set of correlation functions in a given SCFT. From their existence, one can extract useful informations regarding the quantum theory, like OPE coefficients, bounds on numerical factors involved in the bootstrap technique, coefficients of Witten diagrams in the AdS duals, or the computation of exact quantities interpolating between strong and weak couplings regimes. A prototypical example of the topological sector appears in ${\cal N}=4$ SYM in four dimensions \cite{Drukker:2007qr,Pestun:2009nn,Giombi:2009ds}. The dynamics of a particular subset of chiral primary operators and Wilson loops, living on the same $S^2$ embedded in the full space-time, is completely controlled by the zero-instanton sector of the 2D Yang-Mills theory \cite{Bassetto:1998sr}. All the correlation functions do not depend on space-time positions and can be computed in terms of (multi)matrix models \cite{Giombi:2012ep}. The existence of these sectors has been recently generalized to ${\cal N}=4$ SYM in the presence of interface defects \cite{Wang:2020seq}.

In three dimensions, general properties of the superconformal algebra suggest that SCFTs with ${\cal N}\geq4$ always contain a topological sector 
\cite{Chester:2014mea, Beem:2016cbd}. In the ${\cal N}=4$ case, a one-dimensional topological sector has been explicitly constructed in \cite{Dedushenko:2016jxl} as a family of twisted Higgs branch operators belonging to the cohomology of a BRST-like supercharge. Correlation functions of these operators on a line do not depend on the relative separation between the insertion points. As proved there, the cohomological supercharge can be used to  perform supersymmetric localization in a large class of ${\cal N}=4$ theories place on $S^3$. The result is a Matrix Model for a topological quantum mechanics representing the topological sector effectively. All the correlation functions can thus be computed in terms of matrix-integrals. The construction has been later extended to Coulomb branch operators \cite{Dedushenko:2017avn}, complicated by the presence of monopole operators, and to more general manifolds \cite{Panerai:2020boq}. A mini-bootstrap approach has been performed in this sector \cite{Chang:2019dzt}, leading to analytical bounds on flavor central charges and other OPE coefficients. 

The existence of a one-dimensional topological sector finds interesting applications also in the study of ${\cal N}=8$ and ${\cal N}=6$ three-dimensional theories. Here the situation becomes even more interesting since, due to the enhanced supersymmetry the correlation functions of dimension-one topological operators can be related to the ones of the stress-energy tensor in a particular kinematic configuration. The topological sector has thus played a notable role in performing a precision study of maximally supersymmetric (${\cal N} = 8$) SCFTs through conformal bootstrap, allowing to compute exactly some OPE data and constraining "islands" in the parameter space \cite{Chester:2014mea,Agmon:2017xes,Agmon:2019imm}. At the same time, it has been instrumental in fixing contributions to the scattering amplitudes of super-gravitons in M-theory in eleven dimensions \cite{Chester:2018aca}. 

More recently the topological sector of ${\cal N} = 6$ ABJ(M) theory has been also considered in connection with string theory amplitudes in AdS$_4\times {\rm CP}^3$ \cite{Binder:2019mpb}. As already stressed, from a physical point of view ABJ(M) theory is less rigid than its ${\cal N} =8$ cousin. It  admits a variety of limits in which one can compare computations done with different tools and combine results from complementary approaches. On the other hand, the absence of a  ${\cal N} =4$ SYM mirror theory and the presence of Chern-Simons terms have somehow precluded a direct derivation of a one-dimensional action for the topological sector\footnote{We acknowledge Itamar Yaakov for illuminating discussions on this point.}. It then follows that some (reasonable) assumptions made in \cite{Binder:2019mpb} need further support through the use of alternative approaches. More ambitiously, in checking these assumptions one might hope to grasp some hints about the possibility to localize ABJ(M) theory with a supercharge different from the usual KYW one \cite{Kapustin:2009kz}, notably with the supercharge defining the topological sector of the theory.

In this paper, we take a closer look at the topological line of ${\cal N} = 6$ $U(N_1)_k\times U(N_2)_{-k}$ ABJ(M) theory and study the relation between correlation functions of dimension-one topological operators and the mass-deformed Matrix Model of the ABJ(M) parent theory. 
 
As a first step, we present the explicit construction of the one-dimensional topological sector of the ABJ(M) theory obtained by twisting local operators localized on a straight-line parallel to the $x^3$-direction. The superconformal algebra preserved by this line is given by a $\mathfrak{su}(1,1|3)\oplus\mathfrak{u}(1)_b$ inside the original $\mathfrak{osp}(6|4)$ \cite{Bianchi:2017ozk}. We obtain the relevant cohomology working directly in the $\mathfrak{su}(1,1|3)$ formalism. The topological operators are the superconformal primaries of some short irreducible representations of $\mathfrak{su}(1,1|3)$. They are realized explicitly as composite operators of the fundamental matter fields of the theory.  

Focusing on dimension-one topological operators belonging to the stress-energy multiplet, we compute their two-, three- and four-point functions at large $k$ and finite $N_1, N_2$, exploiting standard perturbation theory. First of all, we find a non-vanishing correction to the two-point function at two loops, which turns out to be position independent, thus providing the first confirmation of the topological character of these correlators at quantum level. We then compare these results with a weak coupling expansion of the mass-deformed Matrix Model of the ABJ(M) theory on $S^3$ \cite{Kapustin:2010xq,Jafferis:2010un,Hama:2010av}. 
We find perfect matching between our perturbative results integrated on $S^1 \subset S^3$ and mass derivatives of the mass-deformed Matrix Model. This is a non-trivial confirmation of the assumption made in \cite{Binder:2019mpb} according to which three-dimensional integrated correlators arising in this procedure can be replaced by integrated correlators of the topological operators along the circle, or, in the conformally equivalent set up, along the line in $\mathbb R^3$. Since the validity of this result is a clear indication that an alternative localization procedure should exist for these topological correlators \cite{Agmon:2017xes}, our result provides a first quantitative hint that also in ${\cal N}=6$ case it should be possible to find a one-dimensional theory describing these topological correlators from which one could reconstruct the 3D partition function. In principle, a full-fledged localization computation should provide a construction for such a topological quantum mechanics. Unfortunately, at the moment, there is no such a description for ABJ(M) for $k> 1$.  

Superconformal Ward identities relate the two-point function of dimension-one topological operators to the central charge $c_T$ of the 3D theory. Therefore, as a by-product, from our perturbative calculation we obtain the novel result for the weak coupling expansion of the ABJ(M) central charge up to two loops at generically finite $N_1$ and $N_2$ (see eq. \eqref{centralcharge3}). Remarkably, it coincides with what we obtain from the Matrix Model expansion at weak coupling by applying the prescription in \cite{Closset:2012vg}. 

The paper is organized as follows. In section \ref{sec:generalities} we review the general construction of the topological sector in ${\cal N}\geq4$ SCFTs and discuss the relation that should hold between integrated topological correlators and derivatives of the partition function of the mass-deformed ABJ(M) theory. Using the twisting procedure, in section \ref{sec:line} we explicitly construct the topological operators on the line and obtain their field theory realization. Section \ref{sect: perturbative} is devoted to the perturbative computations of topological correlators at weak coupling, using Feynman diagrams regulated by dimensional reduction. In section \ref{MM}, we present the evaluation of the integrated two-point function and the central charge $c_T$ at weak coupling from the mass-deformed matrix-model, and discuss the matching with the perturbative result at two loops. Exploiting the Matrix Model expansion we also make a prediction for the four-point function at two loops, which results into a non-vanishing, constant contribution that could be checked by evaluating Feynman diagrams as well. Finally, section \ref{conclusions} contains our conclusions and possible new developments. Five appendices follow, which summarise our conventions on the ABJ(M) theory, $\mathfrak{osp}(6|4)$ and $\mathfrak{su}(1,1|3)$ superalgebras, ${\cal N} = 6$ supersymmetry transformations, and provide details on the two-loop calculation.

\section{The topological sector of 3D $\mathcal{N} \geq 4$ theories: A brief review } \label{sec:generalities}

We begin with a brief review of some background material concerning the construction of the one-dimensional topological sector of three-dimensional $\mathcal{N}=4$ SCFTs and its relation with the Matrix Model localizing the theory on $S^3$. We then discuss if and how the generalization to $\mathcal{N}>4$ SCFTs works in general, focusing in particular on the present understanding of $\mathcal{N} =6$ ABJ(M) theory.

\vskip 10pt
Three-dimensional $\mathcal{N}=4$ SCFTs admit a one-dimensional  topological sector, that is a set of operators in the cohomology of a twisted superalgebra, whose correlation functions do not depend on the insertion points when we restrict them to sit on a line in $\mathbb{R}^3$ \cite{Chester:2014mea, Beem:2016cbd}\footnote{The only possible dependence is on the order of the insertions.}. 
These operators turn out to be related to superconformal primaries (SCP) ${\cal O}_{a_1 \dots a_n}(\vec{0})$ of the three-dimensional theory, which belong to short multiplets, have scale dimension and R-symmetry quantum number $\Delta=j = n/2 $ and transform in the $(n+1,1)$ of $SU(2) \times SU(2)$ R-symmetry group.

There are many different reasons why the topological sector plays a relevant role in solving the SCFT. One reason is that it represents a simpler sector where to implement the bootstrap program. Another one is that it is strictly connected with the localization procedure used for evaluating the partition function on $S^3$, so leading to far-reaching consequences in terms of solvability of the theory. We are mostly interested in the latter aspect, which we now review briefly.

As discussed in \cite{Dedushenko:2016jxl}, since the result for the partition function is independent of the supercharge used to localize the functional integral, one can think of localizing the ${\mathcal N}=4$ theory on $S^3$ using the nihilpotent supercharge $Q$ which features the one-dimensional  topological sector, rather than the supercharge originally used in \cite{Kapustin:2009kz}. This procedure leads to a different, but equivalent Matrix Model for the ${\mathcal N}=4$ partition function $\mathcal{Z}[S^3]$, which can be interpreted as coming from the gauge sector coupled to a one-dimensional  Gaussian model localized on the great circle $S^1 \subset S^3$. Remarkably, this one-dimensional factor is exactly the contribution from the one-dimensional  topological sector defined by the $Q$--cohomology, corresponding to $\Delta = j = 1$. 

The non-trivial observation is now the following: Deforming the original SCFT by mass parameters $m^a$ and localizing it on $S^3$ leads to a deformed MM which can be computed exactly in the large $N$ limit \cite{Nosaka:2015bhf,Nosaka:2016vqf}. On the other hand, this is equivalent to add to the one-dimensional  Gaussian model mass terms for the fundamental (bosonic and fermionic) fields ${\cal J}^a$, of the form $-4\pi r^2 m^a \int_{-\pi}^{\pi} d\tau \, {\cal J}^a(\tau)$ \cite{Dedushenko:2016jxl}. Therefore,  taking derivatives of the MM on $S^3$ respect to the mass parameters $m^a$ provides integrated correlation functions of topologically twisted operators living on the great circle.  Precisely, the crucial identity reads  \cite{Agmon:2017xes, Binder:2019mpb} 
\begin{equation}
\label{eq:topcorr}
\Big\langle  \int_{-\pi}^{\pi} \hspace{-0.1cm} d\tau_1 \dots \hspace{-0.1cm}  \int_{-\pi}^{\pi}  \hspace{-0.1cm} d\tau_n \, {\cal J}^{a_1}(\tau_1) \dots {\cal J}^{a_n} (\tau_n) \Big\rangle = \frac{1}{(4\pi r^2)^n} \, \frac{1}{{\cal Z}} \, \frac{\partial^n}{\partial m^{a_1} \dots \partial m^{a_n}} {\cal Z}[S^3,m^a] \Big|_{m^a=0}
\end{equation}
where ${\cal Z}[S^3,m^a]$ is the partition function of the deformed theory on $S^3$ and $r$ is the radius of the sphere. Since the topological correlators are position independent, the integrals on the l.h.s. can be trivially performed leading to a constant factor $(2\pi)^n$ times the correlator.  
Therefore, \eqref{eq:topcorr} provides an exact prescription for computing correlators in the one-dimensional  topological sector in terms of the derivatives of the deformed MM of the three-dimensional theory. Read in the opposite direction, it allows to reconstruct the exact partition function of the three-dimensional theory on the sphere once we have solved the one-dimensional topological theory, i.e. we know exactly all its correlators. 

\vskip 10pt

Prescription \eqref{eq:topcorr} is valid also for ${\mathcal N=8}$ SCFTs \cite{Agmon:2017xes}. In fact, these theories can be seen as a subclass of ${\mathcal N}=4$ theories with $\mathfrak{so}(4)$ flavor symmetry. It is then simply a question of decomposing representations of the ${\mathcal N}=8$ superconformal algebra in terms of the ones of the  ${\mathcal N}=4$ algebra and find the corresponding one-dimensional  topological sector. 
In this case the line operators ${\mathcal J}^a$ come from three-dimensional operators which belong to the ${\cal N}=8$ stress-energy tensor multiplet. Consequently, superconformal Ward identities relate their two-point function $\langle {\cal J}^{a_1}(\tau) {\cal J}^{a_2}(0) \rangle$ to the two-point function of the stress-energy tensor $T_{\mu\nu}$
\begin{equation}\label{stresstensor}
\langle T_{\mu\nu}(\vec{x})T_{\rho\sigma}(0) \rangle=\frac{c_T}{64}(P_{\mu\rho}P_{\nu\sigma}+P_{\nu\rho}P_{\mu\sigma}-P_{\mu\nu}P_{\rho\sigma})\frac{1}{16\pi^2 \vec{x}^{\,2}} 
\end{equation}
where $P_{\mu\nu}=\eta_{\mu\nu}\nabla^2-\partial_\mu\partial_\nu$ and $c_T$ is the central charge of the three-dimensional theory \footnote{\label{fn4} We conventionally set $c_T=1$ both for a real scalar field and a Majorana fermion.}. In particular, one then obtains that 
$c_T$ equals $\langle {\cal J}^{a_1}(\tau) {\cal J}^{a_2}(0) \rangle$, or $\frac{1}{(2\pi)^2} \langle \int{\cal J}^{a_1}(\tau) {\cal J}^{a_2}(0) \rangle$, up to a numerical factor. 

On the other hand, as proved in \cite{Closset:2012vg}, $c_T$ can be independently computed from the mass deformed Matrix Model on $S^3$ as\footnote{We recall that an alternative prescription, which holds for any $\mathcal{N} \geq2$  SCFT, amounts to placing the theory on the squashed sphere $S^3_b$, where $b$ is the squashing parameter. It then follows that the central charge is given by the second derivative of the free energy in the squashed background w.r.t. to $b$ \cite{Closset:2012ru}.}
\begin{equation} \label{centralcharge}
c_T  = - \frac{64}{\pi^2} \, \frac{d^2}{d m^2} \! \log{\cal Z}[S^3,m] \Big|_{m=0} 
\end{equation} 
Therefore, the consistency of the two independent results for $c_T$ -- the one obtained from the topological correlator and the one from \eqref{centralcharge} -- represents an alternative way to prove the validity of \eqref{eq:topcorr}, at least for $n=2$. For the ${\mathcal N}=8$ theories this has been discussed in details in \cite{Agmon:2017xes}.

\vskip 10pt

We are interested in investigating the previous construction for the $\mathcal{N}=6$ $U(N_1)\times U(N_2)$ ABJ(M) theory. Although we should expect things to work similarly, once we decompose $\mathcal{N}=6$ representations in terms on $\mathcal{N}=4$ ones, a rigorous proof of the validity of identity \eqref{eq:topcorr} is still lacking due to the absence of an off-shell formulation of the Chern-Simons sector. 

In \cite{Binder:2019mpb}, assuming that the above derivation holds also for ABJ(M) theory, prescription \eqref{eq:topcorr} has been exploited to fix some coefficients in the Witten diagrams computing four-point functions of topological operators at strong coupling.  
Precisely, describing the ABJ(M) field content in $\mathcal{N}=2$ language, one can turn on a mass deformation in the Matrix Model corresponding to a real mass  spectrum  $(m_+, -m_+, m_-, -m_-)$ for the bifundamental chiral multiplets $({\cal W}_1, \bar{{\cal Z}}_1, {\cal W}_2, \bar{{\cal Z}}_2) \equiv W_{I=1, 2,3,4}$. 
It follows that derivatives of the Matrix Model with respect to $m_{\pm}$ provide integrated correlation functions for the superprimary operators sitting in the stress-energy tensor multiplet (for simplicity we set fermions to zero and consider only the bosonic operators)
\begin{equation}\label{Coperators} 
\mathcal{O}_I^{\; \, J}(\vec{x})= \Tr(C_I(\vec{x}) \bar{C}^J \! (\vec{x})) - \frac14 \delta_I^{\; \,J}  \, \Tr(C_K(\vec{x}) \bar{C}^K \!(\vec{x})) 
\end{equation}
where $C_I$ is the scalar component of $W_I$. 

As for the $\mathcal{N}=8$ case, superconformal Ward identities relate the two-point functions of these operators to correlator \eqref{stresstensor} of the stress-energy tensor. Precisely, we have
\begin{equation}\label{eq:scalarcorr}
\langle \mathcal{O}_I^{\; \, J}(\vec{x}) \, \mathcal{O}_K^{\; \, L}(\vec{0}) \rangle = \frac{c_T }{16} \left( \delta_I^L \delta_K^J - \frac14 \delta_I^J \delta_K^L \right) \, \frac{1}{16 \pi^2 \vec{x}^{\, 2}}
\end{equation}
Assuming that we can still define a topological sector of scalar operators ${\mathcal O}(\tau)$ related to \eqref{Coperators} and localized on the great circle $S^1 \subset S^3$, exploiting \eqref{eq:scalarcorr} we can compute $c_T$ from their two-point function $\langle \mathcal{O}(\tau)\mathcal{O}(0) \rangle$ integrated on $S^1$. 
On the other hand, equation \eqref{centralcharge} is valid also for the ABJ(M) theory in the form 
\begin{equation}\label{eq1.1}
c_T=-\left.\frac{64}{\pi^2} \, \frac{\partial^2}{\partial m^2_\pm} \! \log{{\cal Z}[S^3,m_\pm]} \right|_{m_\pm=0} 
\end{equation}
and provides an alternative way to compute the central charge. Now, if the two results -- the one from the topological correlator and the one from the derivatives of the three-dimensional partition function -- match, we can conclude that \eqref{eq:topcorr} is valid also in the ABJ(M) case. 

This is what we are going to investigate in the rest of the paper.  
After the construction of the topological line operators ${\mathcal O}$, we will check the validity of the following identity\footnote{For notational convenience, in the rest of the paper we choose the radius of the sphere to be $r=1/2$.}
\begin{equation}\label{identity}
 \bigg\langle \int_{-\pi}^\pi \! d\tau_1 \mathcal O(\tau_1) \; \int_{-\pi}^\pi \! d\tau_2  \mathcal O(\tau_2) \bigg\rangle = \frac{1}{\pi^2} \; \frac{\partial^2}{\partial m_{\pm}^2} \! \log{ {\cal Z}[S^3,m^\pm]}  \bigg|_{m_{\pm}=0}
\end{equation}
by matching the weak coupling expansion of the derivatives of the mass deformed ABJ(M) Matrix Model on the r.h.s. against a genuine two-loop calculation of the two-point correlator  $\langle \mathcal{O}(\tau_1)\mathcal{O}(\tau_2) \rangle$. As already mentioned, expressions \eqref{identity} coincide with $- \,
64 c_T$. Therefore, as a by-product, we obtain the central charge of ABJ(M) at weak coupling, up to two-loop order.

\section{The topological line in ABJ(M) theory}\label{sec:line}

This section is devoted to building the {\it topological} sector of local operators in the ABJ(M) theory associated with a straight-line parallel to the $x^3$-direction and parametrized as  $x^\mu(s) = (0,0,s)$, with $s \in (-\infty, + \infty) $ being its proper time. 

The superconformal algebra preserved by this line is given by a $\mathfrak{su}(1,1|3)\oplus\mathfrak{u}(1)_b$ inside the original $\mathfrak{osp}(6|4)$. Our conventions and the commutations relations for these superalgebras are spelled out in appendices \ref{osp64} and \ref{sect: su(1,1|3)}. In the latter, we also clarify our choice of the embedding for the preserved superalgebra inside $\mathfrak{osp}(6|4)$.

\noindent
When constructing this topological sector, we find it convenient to reorganize the scalars $C_I, \bar C^I$ and the fermions $\psi_{I}, \bar\psi^I$, $I=1,2,3,4$,
in irreducible representations of $SU(3)$, the residual R-symmetry group. Precisely, we split them as
\begin{equation}\label{su3breaking}
 C_I=(Z,Y_a) \ \qquad \bar C^I=(\bar Z, \bar Y^a) \ \qquad \psi_{I}=(\psi, \chi_{a}) \ \qquad \bar\psi^I=(\bar\psi, \bar\chi^a) \qquad a = 1,2,3
\end{equation}
where $Y_a (\bar Y^a), \chi_a (\bar\chi^a)$ belong to the ${\mathbf 3} (\bar{\mathbf 3})$ of $SU(3)$, while $Z,\bar Z, \psi, \bar\psi$ are $SU(3)$-singlets.  
Gauge fields split according to the new spacetime symmetry as
\begin{equation}\label{gauge}
A_\mu=(A \equiv A_1-iA_2,\ \bar A \equiv A_1+iA_2,\ A_3)\qquad \hat A_\mu=(\hat A \equiv \hat A_1-i\hat A_2,\ \hat{\bar A} \equiv \hat A_1+i\hat A_2,\ \hat A_3)\qquad   
\end{equation}
together with the corresponding covariant derivatives (see their definition in \eqref{covd}) 
\begin{equation}
D_\mu=(D \equiv D_1-iD_2,\ \bar D \equiv D_1+iD_2,\ D_3)
\end{equation}
\subsection{The Topological Twist}
Recently, the topological twist \cite{Witten:1988ze} has been exploited for constructing two-dimensional protected sectors of four-dimensional $\mathcal{N}\geq 2$ superconformal field theories \cite{Beem:2013sza} and one-dimen\-sional topological sectors of three-dimensional $\mathcal{N}=4,8$ superconformal field theories \cite{Chester:2014mea,Beem:2016cbd}. Below, we use this procedure to single out a topological sector of ABJ(M) theory supported on a line. 

\noindent
The starting point is the complexification of the superalgebra $\mathfrak{su}(1,1|3)$ preserved by the line. Its commutation relations are given in eqs. (\ref{su1,1}, \ref{su3}, \ref{anticomm}, \ref{comm}). Then, inside the complexification of the $\mathfrak{su}(3)$, we can select the $\mathfrak{su}(1,1)(\simeq\mathfrak{sl}(2))$ subalgebra generated by
\begin{equation}\label{sl2}
\mathfrak{su}(1,1)\simeq\bigg\langle i{R_3}^1,\ i{R_1}^3, \ \frac{{R_1}^1-{R_3}^3}{2} \bigg\rangle\equiv\langle {{\cal R}_+, {\cal R}_-, \cal R}_0 \rangle
\end{equation}
These generators obey the following commutation relations
\begin{equation}\label{algebra}
[{\cal R}_0,{\cal R}_\pm]=\pm {\cal R}_\pm\qquad [{\cal R}_+,{\cal R}_-]=-2{\cal R}_0
\end{equation}
We can also define a $\mathfrak{u}(1)$ generator $\frac{{R_1}^1+{R_3}^3}{2}$ that commutes with the  algebra in \eqref{algebra}.

Summarising, we have broken the complexification of the original $\mathfrak{su}(3)$ into $\mathfrak{su}(1,1)\oplus \mathfrak{u}(1)$. 
With respect to this subalgebra, the supercharges split into two
doublets $(Q^1,Q^3)$ and $(S^1,S^3)$, and their hermitian conjugates $(\bar{Q}_1,\bar{Q}_3)$, $(\bar{S}_1,\bar{S}_3)$, which transform in the fundamental of 
$\mathfrak{su}(1,1)$ and have $\mathfrak{u}(1)$ charges $1/6$ and $-1/6$, respectively. The remaining supercharges $Q^2, S^2$ ($\bar{Q}_2, \bar{S}_2$) are instead singlets with $U(1)$ charges $-1/3$ ($1/3$).  

\noindent
The topological twist can now be performed by taking a suitable diagonal sum of the original spacetime conformal algebra defined in \eqref{su1,1} with the $\mathfrak{su}(1,1)$ given in \eqref{sl2}. The twisted generators are
\begin{equation}\label{twist}
\qquad \hat{L}_+=P+{\cal R}_+ \, \qquad \hat{L}_-=K+ {\cal R}_- \, \qquad \hat{L}_0=D+{\cal R}_0
\end{equation}
and satisfy the commutation relations
\begin{equation}\label{algebra1}
[\hat{L}_0,\hat{L}_\pm]=\pm \hat{L}_\pm\qquad [\hat{L}_+,\hat{L}_-]=-2\hat{L}_0
\end{equation}
We shall denote this twisted conformal algebra on the line with $\widehat{\mathfrak{su}}(1,1)$. 

\noindent
Under the new spin assignments induced by $\widehat{\mathfrak{su}}(1,1)$ the supercharges $Q^3$, $S^1$ and their hermitian conjugates are now scalars. In particular, the linear combinations 
\begin{equation}\label{BRST}
\mathcal{Q}_1=Q^{3}+iS^{1}\qquad , \qquad \mathcal{Q}_2=\bar S_3+i \bar Q_1 
\end{equation}
define two independent nihilpotent supercharges, $\mathcal{Q}_1^2=\mathcal{Q}_2^2=0$. Remarkably, the generators of $\widehat{\mathfrak{su}}(1,1)$ are ${\cal Q}$-exact with respect to both charges. In fact, it is easy to check that 
\begin{equation}
\begin{aligned}\label{linealgebra}
\hat{L}_+=\left\{\mathcal{Q}_1, \bar Q_3\right\}=&-i\left\{\mathcal{Q}_2, Q^1\right\} \ \ \ \
\hat{L}_-=-i\left\{\mathcal{Q}_1, \bar S_1\right\}=\left\{\mathcal{Q}_2, S^3\right\} \\
&\hat{L}_0=\frac{1}{2}\left\{\mathcal{Q}_1,\mathcal{Q}_1^\dagger\right\}=\frac{1}{2}\left\{\mathcal{Q}_2,\mathcal{Q}_2^\dagger\right\}
\end{aligned}
\end{equation}
The twisted generators $ \hat{L}_\pm, \hat{L}_0$ and the charges $\mathcal{Q}_1$ and $\mathcal{Q}_2$ span a superalgebra, which possesses a central extension given by
\begin{equation}\label{Z}
\mathcal{Z}=\frac{1}{4}\left\{\mathcal{Q}_1,\mathcal{Q}_2\right\}=\frac{1}{3}M-\frac{{R_3}^3+{R_1}^1}{2}
\end{equation}
where $M$ is the $\mathfrak{u}(1)$ generator defined in \eqref{M}.

\subsection{$\mathcal Q$-cohomology and topological operators}\label{sec:topop}
We now have the necessary ingredients to construct the topological sector of ABJ(M) on the line. It contains all the local\footnote{For our purposes it is sufficient to consider {\em local} operators.}, gauge-invariant operators belonging to the cohomology of a nilpotent charge $\mathcal{Q}$ for which the twisted translations are $\mathcal{Q}-$exact. Since both $\mathcal{Q}_1$ and $\mathcal{Q}_2$
satisfies this property, we can choose either one of them or a suitable linear combination. The results of this section will be independent of which charge we select.

\noindent
The defining conditions for an operator $\mathcal O(s)$ living in the cohomology of $\mathcal Q$ are
$[\mathcal Q, \mathcal O(s)]_\pm=0$, but $\mathcal O(s)\neq [\mathcal Q, \mathcal O'(s)]_\mp$, where either commutators or anticommutators appear depending on the spin of ${\cal O}$. Since $\hat{L}_+$ commutes with ${\mathcal Q}$, we can restrict
our analysis to operators placed at the origin. In fact, an operator at the point $s$ can always be obtained from the one evaluated at the origin by applying a twisted translation, i.e.
\begin{equation}\label{twistedtransl}
\mathcal{O}(s)\equiv e^{is\hat L_+} \, \mathcal O(0) \, e^{-is\hat L_+}
\end{equation}
Moreover, since $\hat L_+$ is ${\cal Q}$-exact, the correlation functions of the operators in the cohomology of ${\cal Q}$ are independent of their position along the line. At most, they can depend on their relative ordering. The operators in the cohomology will be referred to as {\em topological operators}. \\
 Below we focus on solving the constraints
\begin{equation}\label{Ocohom}
[\mathcal Q, \mathcal O(0)]_\pm=0\qquad\text{and}\qquad \mathcal O(0)\neq [\mathcal Q, \mathcal O'(0)]_\mp
\end{equation}
where $\mathcal O(0)$ belongs to a unitary irreducible representations of the superconformal algebra $\mathfrak{su}(1,1|3)$. As briefly reviewed in appendix \ref{irrepline}, the operators in an irreducible representation are classified in terms of  the conformal weight $\Delta$, the $\mathfrak{u}(1)$ charge $m$ and  the eigenvalues $(j_1, j_2)$ corresponding to the two $\mathfrak{su}(3)$ Cartan generators defined in \eqref{su(3)cartan}. We symbolically write $\ket{\Delta, m, j_1, j_2}$ to denote the corresponding state. 

\noindent
The operators solving condition \eqref{Ocohom} can be identified by noting that $\widehat{L}_0$ and $\mathcal Z$, being $\mathcal{Q}$-exact, act trivially within each cohomological class (their action on cohomological representatives is always ${\mathcal Q}$-exact). Therefore, operators obeying the condition \eqref{Ocohom} belong necessarily to the zero eigenspaces of $\hat L_0$ and ${\cal Z}$ \cite{Beem:2013sza}. In particular, in a unitary representation any element of the kernel of $\hat{L}_0$ must be annihilated by ${\cal Q}_1$ and ${\cal Q}_2$, thanks to the last equation in  \eqref{linealgebra}.

\noindent
The problem is then reduced to determining the intersection $\mathcal{N}=\mathrm{Ker}(L_0)\cap \mathrm{Ker}(\mathcal{Z})$. To this end, using the $\mathfrak{su}(3)$ Cartan generators defined in \eqref{su(3)cartan}, we rewrite $\hat L_0$ and ${\cal Z}$ given in eqs. (\ref{twist}, \ref{Z}) as 
$\hat L_0=D-(J_2+J_1)$ and $\mathcal Z=\frac{1}{3}\big(M-(J_2-J_1)\big)$. Therefore, a state $\ket{\Delta, m, j_1, j_2}$ in a given irreducible unitary representation is an eigenvector of $\hat L_0$ and ${\mathcal Z}$ with eigenvalues
\begin{equation}\label{twisteq}
\hat l_0 =\Delta-\frac{j_2+j_1}{2} \; , \qquad \qquad z =\frac{1}{3}\left(m-\frac{j_2-j_1}{2}\right)
\end{equation} 
This state will belong to $\mathcal{N}$ and define a topological operator if and only if 
\begin{equation}\label{finaltwist}
\Delta=\frac{j_2+j_1}{2} \; ,\qquad\qquad m=\frac{j_2-j_1}{2}
\end{equation} 

The next step is to identify these topological operators among the state components of $\mathfrak{su}(1,1|3)$ super-multiplets. 

We begin by scanning the long representations. 
As reviewed in appendix \ref{irrepline}, $\mathcal{A}^{\Delta}_{m;j_1,j_2}$ multiplets are characterized by unitarity constraints \eqref{Amultiplet}. The first of the two possibilities is always incompatible with (\ref{finaltwist}), whereas the second one satisfies (\ref{finaltwist}) at the threshold. Therefore, the superconformal primaries of the $\mathcal A$ multiplets at the threshold certainly belong to the cohomology of $\mathcal{Q}$. However, this identification  can be refined. In fact, due to the recombination phenomenon, the $\mathcal A$ multiplets at the threshold split into short multiplets according to the decomposition \eqref{recomb}. By inspection, we can check that the topological operators are  actually the superprimaries (highest weight operators) of the short multiplets $\mathcal{B}^{\frac{1}{6},\frac{1}{6}}_{\frac{j_2-j_1}{2};j_1,j_2}$ in \eqref{recomb}. This analysis, complemented by the observation that 
no other topological operator arises from the descendants of $\mathcal{B}$ multiplets in decomposition \eqref{recomb}, concludes the identification of topological operators in the $\mathcal{A}^{\Delta}_{m;j_1,j_2}$ class\footnote{More details can be found in \cite{Chester:2014mea,Liendo:2015cgi}.}.

\noindent
Next we look for other candidates by examining the class $\mathcal B$ of short multiplets in more generality. Referring to their shortening conditions (\ref{shortening}, \ref{shortening2}) we immediately see that eqs. \eqref{finaltwist} are always satisfied by the superprimaries of $\mathcal{B}^{\frac{1}{6},0}_{\frac{j_2-j_1}{2};j_1,j_2}$ and $\mathcal{B}^{0,\frac{1}{6}}_{\frac{j_2-j_1}{2};j_1,j_2}$, for generic values of $j_1$ and $j_2$. 

\noindent
Summarizing, we have found that a topological operator is the superprimary of one of the following three multiplets
\begin{equation}
\mathcal{B}^{\frac{1}{6},\frac{1}{6}}_{\frac{j_2-j_1}{2};j_1,j_2}\quad , \qquad\mathcal{B}^{\frac{1}{6},0}_{\frac{j_2-j_1}{2};j_1,j_2}\quad , \qquad \mathcal{B}^{0,\frac{1}{6}}_{\frac{j_2-j_1}{2};j_1,j_2}
\end{equation} 
When $j_1$ or $j_2$ or both vanish these multiplets become even shorter and enhance their supersymmetry. Using the classification reviewed in eqs. (\ref{shortening}--\ref{shortening3}) we can identify them with one of the remaining BPS multiplets.

\subsection{A simple field realization}\label{sect:fr}

The elements of the multiplets determined in the previous subsection can be explicitly realized as composite operators built out of the fundamental matter fields. 
In fact, looking at tables \ref{irrepline}.\ref{table2} and \ref{irrepline}.\ref{table3}, we immediately realise that $Y_1$ and $\bar Y^3$ provide two super-conformal primaries satisfying conditions \eqref{finaltwist} with $(j_1,j_2) = (1,0)$ and $(0,1)$ respectively. Using these two fundamental fields, the simplest gauge-invariant topological operator on the line can be constructed as
\begin{equation}\label{topop0}
\mathcal{O}(s)\equiv e^{is\hat L_+}\mathcal O(0)e^{-is\hat L_+} \qquad {\rm with } \qquad \mathcal O(0)=\Tr(Y_1(0)\bar Y^3(0))  
\end{equation}
and obeys conditions \eqref{finaltwist} with $[\Delta, m, j_1, j_2] = [1,0,1,1]$. 

Evaluating the twisted translation explicitly, at a generic point $s$ on the line this operator can be written as
\begin{equation}\label{topop}
\mathcal{O}(s)=\Tr(Y_a(s)\bar Y^b(s) )\ \bar u^a(s) \, v_b(s) \; , \qquad {\rm with} \quad \; \bar u^a(s) \! = \! (1,0,s) \qquad v_a(s) \! = \! (-s,0,1)
\end{equation}
The contraction with the two polarization vectors leads to a linear combination of single trace operators with coefficients that depend on the insertion points
\begin{equation}\label{topop2}
\mathcal{O}(s)=\Tr(Y_1\bar Y^3 ) - s \Tr(Y_1\bar Y^1 ) + s
\Tr(Y_3\bar Y^3 ) - s^2 \Tr(Y_3\bar Y^1 )
\end{equation}
This is the only topological operator on the line with conformal weight equal to one. 
Generalizations of this operator can be constructed by taking chains of alternating $Y_1$ and $\bar Y^3$ scalars. Gauge invariance forces to have the same number of fields of each type. Therefore, the most general topological operator of this kind, evaluated at $s=0$, reads
\begin{equation}\label{nop}
O_{n}(0)=\Tr(\underbrace{Y_1(0)\bar Y^3(0) \cdots Y_1(0)\bar Y^3(0)}_{n-{\rm times}})
\end{equation}
It satisfies the topological conditions in \eqref{finaltwist} with $[\Delta, m, j_1, j_2] = [n,0,n,n]$. 

These operators exhaust the spectrum of topological, gauge invariant local operators suitable for insertions on the topological line. A larger class of operators can be constructed when one is interested in studying insertions on dynamical defects, like BPS Wilson lines. We will discuss this possibility in a forthcoming paper \cite{forthcoming}.

What makes ${\mathcal O}(0)$ in \eqref{topop0} special within the class of operators \eqref{nop} is that it coincides with the scalar chiral super-primary ${{\mathcal O}_2}^4(0)$ in \eqref{Coperators}, appearing in the super-multiplet of the stress-energy tensor \cite{Binder:2019mpb}. The topological operator in \eqref{topop} is then  ${{\mathcal O}_2}^4$ localized on the line and contracted with the corresponding polarization vectors. As discussed in section \ref{sec:generalities}, it follows that its correlation functions carry some information about the correlation functions of the stress-energy tensor. In particular, 
its two-point function can be used to evaluate the central charge $c_T$ of the theory, as we discuss below.

\vskip 20pt

\section{Topological correlators: The perturbative result}\label{sect: perturbative}

As already mentioned, correlation functions of topological operators \eqref{topop} are expected to be independent of the location of the operators along the line. 
A crucial check of this property comes from the perturbative evaluation of correlators. In particular, whether the topological nature is preserved at the quantum level is one of the main questions that can be addressed within this approach. In fact, if the quantum operator is topological, the evaluation of a generic $n$-point correlator will result in a function whose non-trivial dependence is at most on the coupling constants of the theory. 

Moreover, as discussed in section \ref{sec:generalities} topological correlators are potentially connected with derivatives of the mass-deformed Matrix Model. A confirmation of this intuition comes from proving identity \eqref{identity} perturbatively. 

Motivated by these observations, we study two-, three- and four-point functions. We focus only on {\em connected} correlation functions. While three- and four-point correlators are evaluated up to one loop, we push the calculation for the two-point function up to two loops to provide a check of \eqref{identity} at a non-trivial perturbative order. Correlators are computed on the straight line and later mapped to the great circle in $S^3$, in order to allow for a comparison with localization results discussed in section \ref{MM}.

\subsection{Correlators on the line}
The perturbative evaluation of $n$-point correlation functions relies on the expansion of the Euclidean path integral 
\beq
\langle O(s_1) \cdots O(s_n) \rangle = \int O(s_1) \cdots O(s_n) \, e^{-S}
\eeq
in powers of the coupling constants $N_1/k$ and $N_2/k$. Here $S$ is the ABJ(M) action defined in eqs. (\ref{action}-\ref{S4pt}).  

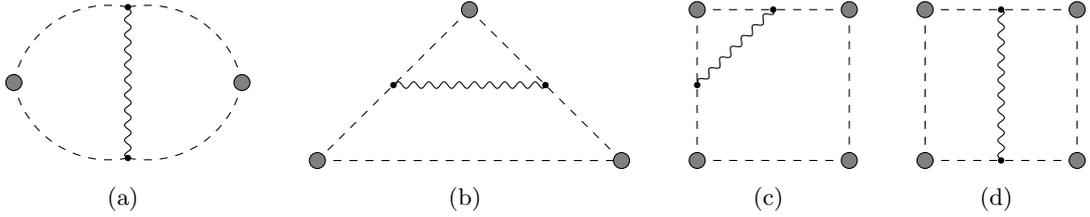
\begin{figure}
\begin{center}
	\subfigure[]{\begin{tikzpicture} \begin{feynman}
		\vertex (a); 
		\vertex[right=1.5cm of a] (b) ;
		\vertex[right=1.5cm of b] (c) ; 
		\vertex[above=1cm of b] (d) ;
		\vertex[below=1cm of b] (e); 
		\diagram* { (a) -- [scalar, quarter left] (d) -- [scalar, quarter left] (c) -- [scalar, quarter left] (e) -- [scalar, quarter left] (a), (d) -- [photon] (e)};
		\draw[fill=gray] (a) circle (3pt);
		\draw[fill=gray] (c) circle (3pt);
		\draw[fill=black] (d) circle (1pt);
		\draw[fill=black] (e) circle (1pt);
		\end{feynman} \end{tikzpicture}\label{fig:phexch}}\qquad
	\subfigure[]{\begin{tikzpicture} \begin{feynman}
		\vertex (a); 
		\vertex[right=1cm of a] (b) ;
		\vertex[right=1cm of b] (c) ;
		\vertex[right=1cm of c] (d) ;		 
		\vertex[right=1cm of d] (e) ;
		\vertex[below=1cm of a] (f); 
		\vertex[below=1cm of e] (g);
		\vertex[above=1cm of c] (h);  
		\diagram* { (f) -- [scalar] (g) -- [scalar] (h) -- [scalar] (f) , (b) -- [photon] (d)};
		\draw[fill=gray] (f) circle (3pt);
		\draw[fill=gray] (g) circle (3pt);
		\draw[fill=gray] (h) circle (3pt);
		\draw[fill=black] (b) circle (1pt);
		\draw[fill=black] (d) circle (1pt);
		\end{feynman} \end{tikzpicture}\label{phexc3}}\qquad
	\subfigure[]{\begin{tikzpicture} \begin{feynman}
		\vertex (a); 
		\vertex[right=2cm of a] (b) ;
		\vertex[above=1cm of a] (e) ;
		\vertex[above=2cm of b] (c) ;
		\vertex[above=2cm of a] (d) ;
		\vertex[right=1cm of d] (f) ;
		\diagram* { (a) -- [scalar] (b) -- [scalar] (c)--[scalar] (d) -- [scalar] (a), (e) -- [photon] (f)};
		\draw[fill=gray] (a) circle (3pt);
		\draw[fill=gray] (b) circle (3pt);
		\draw[fill=gray] (c) circle (3pt);
		\draw[fill=gray] (d) circle (3pt);
		\draw[fill=black] (e) circle (1pt);
		\draw[fill=black] (f) circle (1pt);
		\end{feynman} \end{tikzpicture}\label{tree4pt1}}\qquad
	\subfigure[]{\begin{tikzpicture} \begin{feynman}
		\vertex (a); 
		\vertex[right=2cm of a] (b) ;
		\vertex[right=1cm of a] (e) ;
		\vertex[above=2cm of b] (c) ;
		\vertex[above=2cm of a] (d) ;
		\vertex[right=1cm of d] (f) ;
		\diagram* { (a) -- [scalar] (b) -- [scalar] (c)--[scalar] (d) -- [scalar] (a), (e) -- [photon] (f)};
		\draw[fill=gray] (a) circle (3pt);
		\draw[fill=gray] (b) circle (3pt);
		\draw[fill=gray] (c) circle (3pt);
		\draw[fill=gray] (d) circle (3pt);
		\draw[fill=black] (e) circle (1pt);
		\draw[fill=black] (f) circle (1pt);
		\end{feynman} \end{tikzpicture}\label{quel}}\qquad
	\caption{Topologies of one-loop diagrams contributing to the correlators.}
	\label{fig:1loop3a4}
\end{center}
\end{figure}

Performing all possible contractions and using the scalar propagator in \eqref{scalartree}, for tree-level connected correlators we obtain
\begin{equation}
\braket{\Op(s)\Op(0)}^{(0)}=  \bar u^a(s) \, v_b(s) \, \braket{\Tr(Y_a\bar Y^b ) \Tr(Y_1\bar Y^3)}^{(0)} = - \frac{N_1 N_2}{(4 \pi)^2}
\end{equation}
\beq
\hspace{-0.5cm} \braket{\Op(t) \Op(s)\Op(0)}^{(0)} =  \bar u^a(t) v_b(t) \bar u^c(s) v_d(s) \braket{\Tr(Y_a\bar Y^b ) \Tr(Y_c\bar Y^d ) \Tr(Y_1\bar Y^3)}^{(0)} \; = \; 0
\eeq
\bea\label{4pt}
&& \braket{\Op(z) \Op(t) \Op(s)\Op(0)}^{(0)} =  \\
&=& \bar u^a(z) v_b(z) \bar u^c(t) v_d(t) \bar u^e(s) v_f(s) \braket{\Tr(Y_a\bar Y^b ) \Tr(Y_c\bar Y^d ) \Tr(Y_e\bar Y^f )\Tr(Y_1\bar Y^3)}^{(0)} \; = \; 2\frac{N_1 N_2}{(4\pi)^4} \nonumber
\eea

In the non-vanishing cases the worldline dependence at the denominator encoded in the propagators is canceled by an analogous numerator coming from the contraction of the polarization vectors. 

\vskip 10pt
One-loop corrections to two-, three- and four-point functions are drawn in figure \ref{fig:1loop3a4}. It is easy to realize that they all vanish due to geometrical reasons. All the contributions are proportional to one Levi-Civita tensor $\varepsilon_{\mu\nu\rho}$ coming from the gauge propagator (see eq. \eqref{prop:vector}), which is contracted with spacetime derivatives coming from either internal vertices or the gauge propagator. It is a matter of the fact that such structures eventually vanish when projected on the line. 

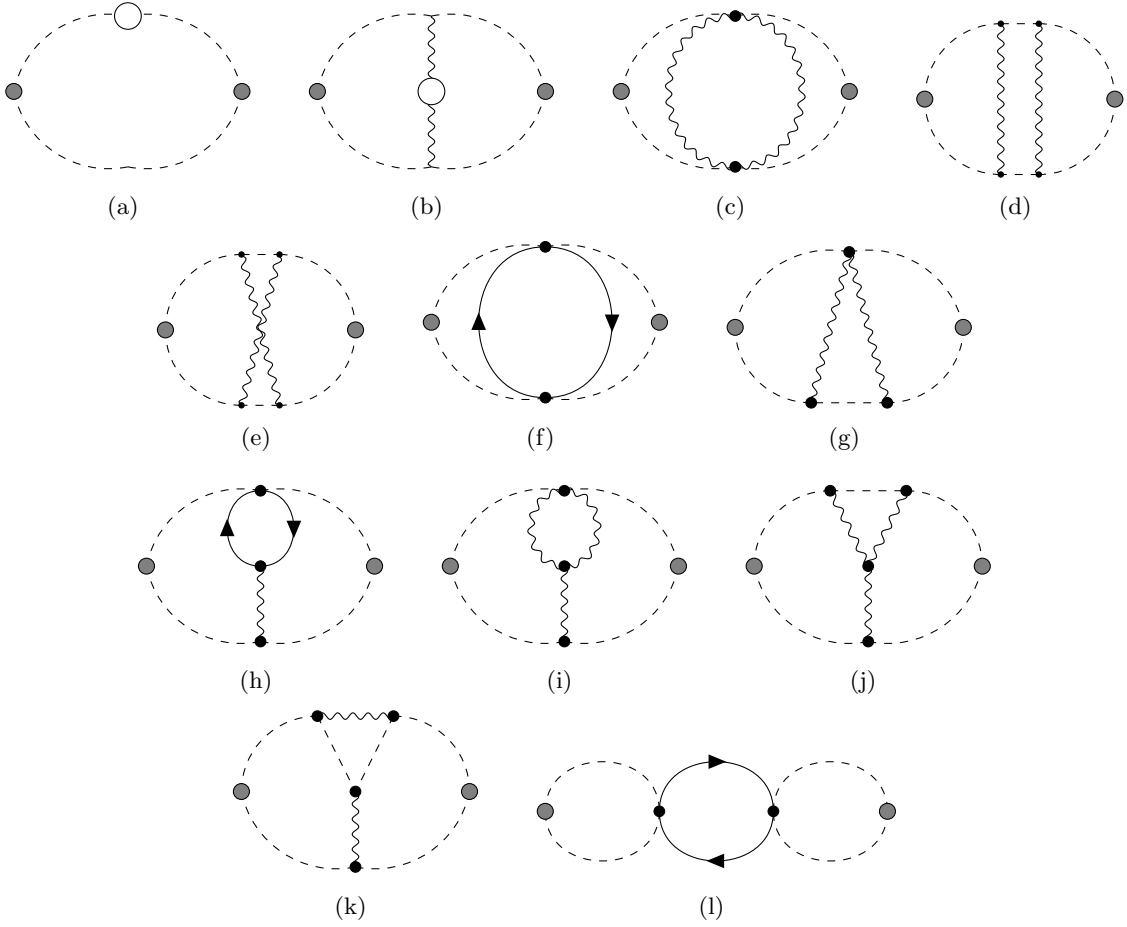
\begin{figure}[h]
	\centering
\subfigure[]{\begin{tikzpicture} \begin{feynman}
		\vertex (a); 
		\vertex[right=1.5cm of a] (b) ;
		\vertex[right=1.5cm of b] (c) ; 
		\vertex[above=1cm of b] (d) ;
		\vertex[below=1cm of b] (e); 
		\diagram* { (a) -- [scalar, quarter left] (d) -- [scalar, quarter left] (c) -- [scalar, quarter left] (e) -- [scalar, quarter left] (a)};
		\draw[fill=gray] (a) circle (3pt) ;
		\draw[fill=white] (d) circle (5pt) ;
		\draw[fill=gray] (c) circle (3pt);
		\end{feynman} \end{tikzpicture}\label{scalarcorr}}\qquad
\subfigure[]{\begin{tikzpicture} \begin{feynman}
		\vertex (a); 
		\vertex[right=1.5cm of a] (b) ;
		\vertex[right=1.5cm of b] (c) ; 
		\vertex[above=1cm of b] (d) ;
		\vertex[below=1cm of b] (e); 
		\diagram* { (a) -- [scalar, quarter left] (d) -- [scalar, quarter left] (c) -- [scalar, quarter left] (e) -- [scalar, quarter left] (a), (d) -- [photon] (b) -- [photon] (e)};
		\draw[fill=gray] (a) circle (3pt);
		\draw[fill=white] (b) circle (5pt);
		\draw[fill=gray] (c) circle (3pt);
		\end{feynman} \end{tikzpicture}\label{kite}}\qquad
\subfigure[]{\begin{tikzpicture} \begin{feynman}
		\vertex (a); 
		\vertex[right=1.5cm of a] (b) ;
		\vertex[right=1.5cm of b] (c) ; 
		\vertex[above=1cm of b] (d) ;
		\vertex[below=1cm of b] (e); 
		\diagram* { (a) -- [scalar, quarter left] (d) -- [scalar, quarter left] (c) -- [scalar, quarter left] (e) -- [scalar, quarter left] (a), (d) -- [photon,half left] (e) -- [photon, half left] (d)};
		\draw[fill=gray] (a) circle (3pt);
		\draw[fill=gray] (c) circle (3pt);
		\draw[fill=black] (d) circle (2pt);
		\draw[fill=black] (e) circle (2pt);
		\end{feynman} \end{tikzpicture}\label{doublegauge}}\qquad
\subfigure[]{\begin{tikzpicture} \begin{feynman}
		\vertex (a); 
		\vertex[right=1cm of a] (b) ;
		\vertex[right=.5cm of b] (c) ; 
		\vertex[right=1cm of c] (f) ;
		\vertex[above=1cm of b] (d) ;
		\vertex[below=1cm of b] (e); 
		\vertex[above=1cm of c] (g) ;
		\vertex[below=1cm of c] (h); 
		\diagram* { (a) -- [scalar, quarter left] (d) -- [scalar] (g) -- [scalar, quarter left] (f) -- [scalar, quarter left] (h) -- [scalar] (e) -- [scalar,quarter left] (a), (d) -- [photon] (e), (g) -- [photon] (h)};
		\draw[fill=gray] (a) circle (3pt);
		\draw[fill=gray] (f) circle (3pt);
		\draw[fill=black] (d) circle (1pt);
		\draw[fill=black] (e) circle (1pt);
		\draw[fill=black] (g) circle (1pt);
		\draw[fill=black] (h) circle (1pt);
		\end{feynman} \end{tikzpicture}\label{duobleph}}\qquad
\subfigure[]{\begin{tikzpicture} \begin{feynman}
		\vertex (a); 
		\vertex[right=1cm of a] (b) ;
		\vertex[right=.5cm of b] (c) ; 
		\vertex[right=1cm of c] (f) ;
		\vertex[above=1cm of b] (d) ;
		\vertex[below=1cm of b] (e); 
		\vertex[above=1cm of c] (g) ;
		\vertex[below=1cm of c] (h); 
		\diagram* { (a) -- [scalar, quarter left] (d) -- [scalar] (g) -- [scalar, quarter left] (f) -- [scalar, quarter left] (h) -- [scalar] (e) -- [scalar,quarter left] (a), (g) -- [photon] (e), (d) -- [photon] (h)};
		\draw[fill=gray] (a) circle (3pt);
		\draw[fill=gray] (f) circle (3pt);
		\draw[fill=black] (d) circle (1pt);
		\draw[fill=black] (e) circle (1pt);
		\draw[fill=black] (g) circle (1pt);
		\draw[fill=black] (h) circle (1pt);
		\end{feynman} \end{tikzpicture}\label{crossedduobleph}}\qquad
\subfigure[]{\begin{tikzpicture} \begin{feynman}
		\vertex (a); 
		\vertex[right=1.5cm of a] (b) ;
		\vertex[right=1.5cm of b] (c) ; 
		\vertex[above=1cm of b] (d) ;
		\vertex[below=1cm of b] (e); 
		\diagram* { (a) -- [scalar, quarter left] (d) -- [scalar, quarter left] (c) -- [scalar, quarter left] (e) -- [scalar, quarter left] (a), (d) -- [fermion,half left] (e) -- [fermion, half left] (d)};
		\draw[fill=gray] (a) circle (3pt);
		\draw[fill=gray] (c) circle (3pt);
		\draw[fill=black] (d) circle (2pt);
		\draw[fill=black] (e) circle (2pt);
		\end{feynman} \end{tikzpicture}\label{4ptccff}}\qquad		
\subfigure[]{\begin{tikzpicture} \begin{feynman}
	\vertex (a); 
	\vertex[right=1cm of a] (b) ;
	\vertex[right=.5cm of b] (c) ; 
	\vertex[right=.5cm of c] (d) ;
	\vertex[right=1cm of d] (e); 
	\vertex[below=1cm of b] (f);
	\vertex[above=1cm of c] (h);
	\vertex[below=1cm of d] (g);  
	\diagram* { (a) -- [scalar, quarter left] (h) -- [scalar, quarter left] (e) -- [scalar, quarter left] (g) --[scalar] (f) -- [scalar, quarter left] (a), (h) -- [photon] (f), (h) -- [photon] (g)};
	\draw[fill=gray] (a) circle (3pt);
	\draw[fill=gray] (e) circle (3pt);
	\draw[fill=black] (f) circle (2pt);
	\draw[fill=black] (g) circle (2pt);
	\draw[fill=black] (h) circle (2pt);
	\end{feynman} \end{tikzpicture}\label{triangle}}\qquad
	
\subfigure[]{\begin{tikzpicture} \begin{feynman}
		\vertex (a); 
		\vertex[right=1.5cm of a] (b) ;
		\vertex[right=1.5cm of b] (c) ; 
		\vertex[above=1cm of b] (d) ;
		\vertex[below=1cm of b] (e); 
		\diagram* { (a) -- [scalar, quarter left] (d) -- [scalar, quarter left] (c) -- [scalar, quarter left] (e) -- [scalar, quarter left] (a), (d) -- [fermion,half left] (b) -- [fermion, half left] (d), (b) -- [photon] (e)};
		\draw[fill=gray] (a) circle (3pt);
		\draw[fill=gray] (c) circle (3pt);
		\draw[fill=black] (d) circle (2pt);
		\draw[fill=black] (e) circle (2pt);
		\draw[fill=black] (b) circle (2pt);
		\end{feynman} \end{tikzpicture}\label{fermgauge}}\qquad
\subfigure[]{\begin{tikzpicture} \begin{feynman}
	\vertex (a); 
	\vertex[right=1.5cm of a] (b) ;
	\vertex[right=1.5cm of b] (c) ; 
	\vertex[above=1cm of b] (d) ;
	\vertex[below=1cm of b] (e); 
	\diagram* { (a) -- [scalar, quarter left] (d) -- [scalar, quarter left] (c) -- [scalar, quarter left] (e) -- [scalar, quarter left] (a), (d) -- [photon,half left] (b) -- [photon, half left] (d), (b) -- [photon] (e)};
	\draw[fill=gray] (a) circle (3pt);
	\draw[fill=gray] (c) circle (3pt);
	\draw[fill=black] (d) circle (2pt);
	\draw[fill=black] (e) circle (2pt);
	\draw[fill=black] (b) circle (2pt);
	\end{feynman} \end{tikzpicture}\label{gaugethree}}\qquad
\subfigure[]{\begin{tikzpicture} \begin{feynman}
	\vertex (a); 
	\vertex[right=1cm of a] (b) ;
	\vertex[right=.5cm of b] (c) ; 
	\vertex[right=.5cm of c] (d) ;
	\vertex[right=1cm of d] (e); 
	\vertex[above=1cm of b] (f);
	\vertex[above=1cm of d] (g);
	\vertex[below=1cm of c] (h);
	\diagram* { (a) -- [scalar, quarter left] (f) -- [scalar] (g) -- [scalar, quarter left] (e) -- [scalar, quarter left] (h) -- [scalar, quarter left] (a), (f) -- [photon] (c), (g) -- [photon] (c) -- [photon] (h)};
	\draw[fill=gray] (a) circle (3pt);
	\draw[fill=gray] (e) circle (3pt);
	\draw[fill=black] (f) circle (2pt);
	\draw[fill=black] (g) circle (2pt);
	\draw[fill=black] (c) circle (2pt);
	\draw[fill=black] (h) circle (2pt);
	\end{feynman} \end{tikzpicture}\label{fork}}\qquad
\subfigure[]{\begin{tikzpicture} \begin{feynman}
	\vertex (a); 
	\vertex[right=1cm of a] (b) ;
	\vertex[right=.5cm of b] (c) ; 
	\vertex[right=.5cm of c] (d) ;
	\vertex[right=1cm of d] (e); 
	\vertex[above=1cm of b] (f);
	\vertex[above=1cm of d] (g);
	\vertex[below=1cm of c] (h);
	\diagram* { (a) -- [scalar, quarter left] (f) -- [photon] (g) -- [scalar, quarter left] (e) -- [scalar, quarter left] (h) -- [scalar, quarter left] (a), (f) -- [scalar] (c), (g) -- [scalar] (c) -- [photon] (h)};
	\draw[fill=gray] (a) circle (3pt);
	\draw[fill=gray] (e) circle (3pt);
	\draw[fill=black] (f) circle (2pt);
	\draw[fill=black] (g) circle (2pt);
	\draw[fill=black] (c) circle (2pt);
	\draw[fill=black] (h) circle (2pt);
	\end{feynman} \end{tikzpicture}\label{scalarfork}}\qquad
	\subfigure[]{\begin{tikzpicture} \begin{feynman}
	\vertex (a); 
	\vertex[right=1.5cm of a] (b) ;
	\vertex[right=1.5cm of b] (c) ; 
	\vertex[right=1.5cm of c] (d) ;
	\diagram* { (a) -- [scalar, half left] (b) -- [scalar, half left] (a),(b) -- [fermion, half left] (c) -- [fermion, half left] (b), (c) -- [scalar, half left] (d) -- [scalar, half left] (c),};
	\draw[fill=gray] (a) circle (3pt);
	\draw[fill=gray] (d) circle (3pt);
	\draw[fill=black] (b) circle (2pt);
	\draw[fill=black] (c) circle (2pt);
	\end{feynman} \end{tikzpicture}\label{threeeyes}}\qquad
	\caption{Two-loop diagrams for the two-point function. In (a) the white circle is the two-loop correction to the scalar propagator, while in (b) the circle is the one-loop correction to the gauge field propagator. Diagrams (h), (i), (j) and (k) sum up to provide the vertex correction.}
\end{figure}

\vskip 10pt
The first non-trivial information comes at two loops. We restrict the evaluation to the two-point function, whose diagrams at this order are given in figures \ref{scalarcorr}-\ref{fork}.

The corresponding algebraic expressions, including the combinatorial and color factors are listed in appendix \ref{app:twoloop}. 
We evaluate the corresponding integrals by Fourier transforming to momentum space. Potential UV divergences are regularized within the DRED scheme \cite{Siegel:1979wq,Siegel:1980qs}. This amounts to first perform the tensor algebra strictly in three dimensions to reduce the integrals to a linear combination of scalar integrals and then analytically continue the resulting integrals to $d=3-2\epsilon$ dimensions. As usual, we also introduce a dimensionful parameter $\mu$ to correct the scale dimensions of the couplings when they are promoted to $d$ dimensions. 

Applying {\em Mathematica} routines\footnote{We are grateful to Marco Bianchi for sharing with us his routines.} based on the uniqueness method the momentum integrals can be analytically evaluated, leading to the results for every single diagram listed below. 

Starting from the first diagram in fig. \ref{scalarcorr} we have
\begin{equation}
\eqref{scalarcorr} = - \frac{\Gamma^3\left(\frac12 - \epsilon\right)}{4^3 \pi^{\frac92 - 3 \epsilon}} \; \frac12 {\cal C}(N_1,N_2) \, |\mu s|^{8\epsilon} 
\end{equation}
where ${\cal C}(N_1,N_2) $ is the two-loop correction to the scalar propagator computed in \cite{Bianchi:2018bke}. Its expansion at small $\epsilon$ is given in eq. \eqref{eq:2loopscalar}. Therefore, neglecting terms which go to zero in the $\epsilon \to 0$ limit, the contribution of this diagram reads
\[
\begin{aligned}
\eqref{scalarcorr} &= \frac{N_1 N_2}{k^2} \, \frac{1}{128 \pi^2} \, \Big[ -\left(N_1^2+N_2^2 + 4 N_1 N_2-6\right) \frac{1}{ \epsilon }\\
&\ \quad + (N_1^2+N_2^2-2) \left(\pi ^2-2 (3+\log 2 )\right)+4 (N_1 N_2-1) \left(\pi ^2-2 (11+\log 2 )\right) \Big] \, |\mu s|^{8\epsilon} 
\end{aligned}
\]
For the rest of the diagrams, neglecting terms that vanish for $\epsilon \to 0$, we obtain
\[
\begin{aligned}
\eqref{kite}&= -\frac{N_1 N_2}{k^2} \, \left(N_1 N_2-1\right) \, \frac{\left(\pi ^2-12\right)}{16 \pi ^2} \, |\mu s|^{8\epsilon}  
\\
\eqref{doublegauge} &= \frac{N_1 N_2}{k^2} \left(N_1^2+N_2^2-4 N_1 N_2+2\right) \left(\frac{1}{128 \pi ^2}\frac{1}{ \epsilon }+\frac{1+\log 2}{64 \pi ^2}\right) \, |\mu s|^{8\epsilon} 
 \\
\eqref{duobleph}&= 0
 \\
\eqref{crossedduobleph} &= -\frac{N_1 N_2}{k^2} \, \left(N_1 N_2-1\right) \, \frac{\left(5 \pi ^2-48\right)}{96 \pi ^2} \, |\mu s|^{8\epsilon} 
 \\
\eqref{4ptccff} &= \frac{N_1 N_2}{k^2} \, \left(N_1 N_2-1\right) \left(\frac{1}{16 \pi ^2}\frac{1}{ \epsilon }+\frac{1+\log 2}{8 \pi ^2}\right) \, |\mu s|^{8\epsilon} 
 \\
\eqref{triangle}&= -\frac{ N_1 N_2}{k^2} \, \left(N_1^2+N_2^2-4 N_1 N_2+2\right) \, \frac{\left(\pi ^2-12\right)}{128 \pi ^2 } \, |\mu s|^{8\epsilon} 
 \\
\eqref{fermgauge}&=0
 \\
\eqref{gaugethree}&=0
 \\
\eqref{fork}&= \frac{ N_1 N_2}{k^2} \, \left(N_1 N_2-1\right) \, \frac{\left(\pi ^2-12\right)}{48 \pi ^2} \, |\mu s|^{8\epsilon} 
 \\
\eqref{scalarfork}&= \frac{N_1 N_2}{k^2} \, \left(N_1^2+N_2^2-2\right) \, \frac{\left(\pi ^2-12\right)}{192 \pi ^2} \, |\mu s|^{8\epsilon} 
 \\
\eqref{threeeyes}&= - \frac{N_1N_2}{k^2} \, \left(N_1-N_2\right)^2 \frac{1}{64} \, |\mu s|^{8\epsilon} 
\end{aligned}
\]

Summing all the contributions, it is easy to realize that the $\epsilon$-poles cancel exactly. We can then safely take the $\epsilon \to 0$ limit and the final result for the two-point function, up to two loops reads
\begin{equation}\label{pertresult}
\langle \mathcal{O}(s) \mathcal{O}(0) \rangle^{(2)} = - \frac{N_1 N_2}{(4\pi)^2} \, \left( 1 - \frac{\pi^2}{6k^2} (N_1^2 + N_2^2 - 2) \right)
\end{equation}
We note that for dimensional reasons all the diagramatic contributions have a dipendence on the position of the form $|\mu s|^{8\epsilon} $. In principle, expanding $|\mu s|^{8\epsilon} \sim (1 + 8 \epsilon \log{|\mu s|} + \cdots)$ might produce dangerous, finite $\log{|\mu s|}$ terms that would spoil the topological nature of the operators at quantum level. However, this does not happen, thanks to the complete cancellation of the $1/\epsilon$ poles. Therefore, the BPS nature that the operators possess in the parent three-dimensional theory nicely protects the correlators on the line, which then turns out to be topological also at quantum level.

\subsection{Correlators on the circle}
If we assume that there are no conformal anomalies at quantum level, correlators of twisted operators computed on a line embedded in $\mathbb R^3$ and on the great circle $S^1 \subset S^3$ should be exactly the same \cite{Dedushenko:2016jxl}. In other words, it is reasonable to assume that (setting $s=\tan{\frac{\tau}{2}}$)
\begin{align}
\label{topcorr} \ev{{\mathcal O}(s_1)\dots {\mathcal O}(s_k)}_{\mathbb{R^3}}&=\ev{{\mathcal O}_S(\tau_1)\dots {\mathcal O}_S(\tau_k)}_{S^3}
\end{align}
where ${\mathcal O}(s)$ is the operator in \eqref{topop} on the line and the operator ${\mathcal O}_S(\tau)$ is its counterpart on the circle obtained by contracting the $S^3$  operator localized on $S^1$ with polarization vectors $\bar{u}_S^a, {v_S}_a$ on the great circle. 

From the background independence of the topological correlators stated in eq. \eqref{topcorr} it is easy to infer how the polarization vectors get mapped from the line to the great circle. In fact, taking into account that 
the ABJ(M) scalar fields transform under a conformal transformation as $Y_1 = \Lambda^{\frac12} (Y_S)_1$, $\bar{Y}^3 = \Lambda^{\frac12} \bar{Y}_S^3$, with $\Lambda = \cos^2{\frac{\tau}{2}}$ being the conformal factor, from the cohomological identification ${\mathcal O}(s) = {\mathcal O}_S(\tau)$ we obtain 
\begin{equation}
\bar{u}_S^a =  \Lambda^{\frac12} \bar{u}^a = \left(\cos\frac{\tau}{2}, 0 , \sin \frac{\tau}{2}\right)
\qquad {v_S}_a = \Lambda^\frac12 v_a = \left(-\sin\frac{\tau}{2},0,\cos\frac{\tau}{2}\right)
\end{equation}
where $\bar{u}^a, v_a$ have been defined in \eqref{topop}. 

\vskip 20pt
\subsection{The central charge at weak coupling}

As discussed in section \ref{sec:topop}, the topological operators in \eqref{topop}  are related to the superprimaries \eqref{Coperators} of the stress-energy tensor localized on the line. In SU(4) notation we can indeed write
\begin{equation}
{\cal O}(s) = {\cal O}_I^J (0,0,s) \bar{U}^I(s) V_J(s) \, ,  \qquad {\rm with} \qquad \bar{U}^I(s) = (0,1,0,s) \, , \quad V_J(s) = (0,-s,0,1)
\end{equation} 
It follows that if we project the two-point function \eqref{eq:scalarcorr} on the line by setting $\vec{x} = (0,0,s)$ and multiplying both sides by $\bar{U}^I(s) V_J(s) \bar{U}^K(0) V_L(0)$, we obtain an explicit expression for the central charge in terms of the topological correlator
\begin{equation}\label{centralcharge2}
c_T = - 64 \, (2\pi)^2 \, \langle {\cal O}(s) \, {\cal O}(0) \rangle = -64 \;  \Big\langle {\int_{-\pi}^\pi \! d\tau_1 \cal O}(\tau_1) \,
\int_{-\pi}^\pi \! d\tau_2 {\cal O}(\tau_2) \Big\rangle 
\end{equation}
where the second equality has been obtained by taking into account the translational invariance of the correlator and the identity $\langle {\cal O}(s) \, {\cal O}(0) \rangle_{\mathbb{R}^3} = \langle {\cal O}(s) \, {\cal O}(0) \rangle_{S^3}$ discussed above.

Inserting in \eqref{centralcharge2} the perturbative result \eqref{pertresult} for the two-point function we obtain the expansion of the ABJ(M) central charge at second order in the couplings and at generic, finite values of the group ranks
\begin{equation}\label{centralcharge3}
c_T = 16N_1N_2\ \Big(1-\frac{\pi^2}{6 k^2}(N_1^2+N_2^2-2)\ +O\left(\frac{1}{k^3}\right)\Big)
\end{equation}
We note that at tree level it reproduces correctly the central for a free theory of $4(N_1 \times N_2)$ chiral multiplets, in agreement with our conventions (see footnote \ref{fn4}), while for $N_1=N_2=2$, we correctly recover the two-loop approximation of $c_T$ in eq. (5.20) of \cite{Chester:2014fya}.

\section{The Matrix Model expansion at weak coupling}\label{MM}

\subsection{The main result}
We are almost ready to prove identity \eqref{identity}. The last ingredient that we need is the weak coupling expansion of the mass-deformed Matrix Model of ABJ(M) on $S^3$ \cite{Kapustin:2010xq,Jafferis:2010un,Hama:2010av} and its second derivatives respect to the masses, to be compared with the perturbative result \eqref{pertresult} for the topological two-point function.

To this end we consider the mass-deformed Matrix Model of the ABJ(M) theory \cite{Kapustin:2010xq,Jafferis:2010un,Hama:2010av}
\begin{equation}\label{Partitionmatrix}
Z=\frac{1}{(N!)^2}\int d\lambda \,d\mu\, \frac{e^{i\pi k\sum_i\left(\lambda_i^2-\mu_i^2\right)} \prod_{i<j}16 \sinh^2\left[\pi\left(\lambda_i-\lambda_j\right)\right]\sinh^2\left[\pi\left(\mu_i-\mu_j\right)\right]}
{\prod_{i,j}4\cosh\left[\pi(\lambda_i-\mu_j)+\frac{\pi m_+}{2}\right]\cosh\left[\pi(\lambda_i-\mu_j)+\frac{\pi m_-}{2}\right]}
\end{equation}
where the mass assignment is the one recalled in section \ref{sec:generalities} \cite{Binder:2019mpb}. Taking derivatives respect to $m_-$ (the same result would rise taking derivatives respect to $m_+$) we immediately find 
\begin{equation}\label{twoderivative}
\frac{\partial^2}{\partial m^2_{-}} \log{Z[S^3, m_\pm]}\bigg|_{m_\pm=0}=\frac{Z''}{Z}-\left(\frac{Z'}{Z}\right)^2    
\end{equation}
where $Z$ is the undeformed MM, whereas its derivatives are given by
\begin{align}
Z'&=-\frac{1}{(N!)^2}\int d\lambda \,d\mu \, e^{i\pi k\sum_i\left(\lambda_i^2-\mu_i^2\right)}\, Z_{1-\textup{loop}}(\lambda_i,\,\mu_j)   \sum_{i,j}\tanh\pi(\lambda_i-\mu_j)\\
Z''&=\frac{1}{(N!)^2}\int d\lambda \,d\mu \, e^{i\pi k\sum_i\left(\lambda_i^2-\mu_i^2\right)}\, Z_{1-\textup{loop}}(\lambda_i,\,\mu_j) \label{secder}\\
&\quad \quad \times \frac{\pi^2}{4}\left( \left(\sum_{i,j}\tanh(\pi(\lambda_i-\mu_j))\right)^2-\sum_{i,j}\frac{1}{\cosh^2(\pi\left(\lambda_i-\mu_j)\right)}
\right)\nonumber
\end{align}
with
\begin{equation}
Z_{1-\textup{loop}}(\lambda_i,\,\mu_j)=\frac{ \prod_{i<j}16 \sinh^2\left[\pi\left(\lambda_i-\lambda_j\right)\right]\sinh^2\left[\pi\left(\mu_i-\mu_j\right)\right]}
{\prod_{i,j}4\cosh(\pi\left(\lambda_i-\mu_j\right))\cosh(\pi\left(\lambda_i-\mu_j\right))}
\end{equation} 
Since the integrand in $Z'$ is odd under $\lambda\leftrightarrow\mu$ exchange, it vanishes once integrated. Thus we only need to compute contribution \eqref{secder}. Performing the following change of variables
\begin{equation}
x_i= \pi\sqrt{k}\lambda_i\,, \qquad y_j=\pi\sqrt{k}\mu_j\,, \qquad g_s=\frac{1}{\sqrt{k}}    
\end{equation}
the relevant quantities become
\begin{align}\label{Zsec}
Z=&\int \! dX\,dY e^{\frac{i}{\pi}\sum_i\left(x_i^2-y_i^2\right)}f(x,y) \\
Z''=&\int \! dX\,dY e^{\frac{i}{\pi}\sum_i\left(x_i^2-y_i^2\right)}f(x,y)\ \frac{\pi^2}{4}\left( \!
 \left(\sum_{i,j}\tanh( g_s(x_i-y_j))\right)^{\! 2} \! - \! \sum_{i,j}
\frac{1}{\cosh^2( g_s\left(x_i-y_j\right))}\right) \nonumber 
\end{align}
where $dX, dY$ are the Haar measures and 
\begin{equation}\label{function}
f(x,y)=\prod_{i<j}\frac{\sinh^2( g_s(x_i-x_j))}{g_s^2(x_i-x_j)^2}\frac{\sinh^2(g_s(y_i-y_j))}{g_s^2(y_i-y_j)^2}\frac{1}{\prod_{i,j}\cosh^2( g_s(x_i-y_j))}
\end{equation}
In order to compute $Z$ and $Z''$, we find it convenient to canonically normalize them as $Z'' \to Z''/Z_0 \equiv  {\mathcal Z''}$, $Z \to Z/Z_0 \equiv {\mathcal Z}$ where 
\begin{equation}\label{Z0}
Z_0 \equiv \int dXdY e^{\frac{i}{\pi}\sum_i\left(x_i^2-y_i^2\right)}
\end{equation}
is the free partition function. By perturbatively expanding the integrands in (\ref{Zsec}) up to $g_s^4 \sim \frac{1}{k^2}$, i.e. at two loops, and evaluating the normalized gaussian matrix integrals, we obtain
\begin{equation}
\begin{aligned}\label{ZZsec}
\mathcal Z''=-\frac{\pi^2}{4}N_1N_2 &\Big[ 1 +g_s^2\ \frac{i\pi}{6} \ (N_2 - N_1) \ \left(1 - (N_2 - N_1)^2 \right)  \\ 
& - g^4_s\ \frac{\pi^2}{72}   \Big(-24 + 16 N_2^2  - 12N_1( N_2 - N_1) + N_2^4 + 6N_2^2 N_1^2 + 2 N_2N_1^3 - N_1^4 \\
&\ \qquad \qquad   + (N_2 - N_1)^6  \Big) +O(g_s^6)\Big] \\\\
\frac{1}{\mathcal Z} =1 - g_s^2\ \frac{i \pi}{6}& \ (N_2 - N_1) \left(1 - (N_2 - N_1)^2 \right) \\
& \hspace{-1.5cm} - g_s^4\  \frac{\pi^2}{72}\Big( -2(N_2^2 - N_1^2) + 8 N_2N_1 - 5 N_2^4 + 2N_2N_1(N_2-N_1)(8N_2 - 7N_1) - 3N_1^4  \\
&\qquad + (N_2 -  N_1)^6   \Big) +O(g_s^6) 
\end{aligned} 
\end{equation}
If we now substitute back $g_s\rightarrow\frac{1}{\sqrt k}$, the final result reads 
\begin{equation}\label{final}
\frac{1}{\pi^2} \, \frac{\partial^2}{\partial m^2_{-}} \log{Z[S^3, m_\pm]}\bigg|_{m_\pm=0}= \frac{1}{\pi^2} \,  \frac{\mathcal{Z''}}{\mathcal  Z}=-\frac{N_1N_2}{4} \\ \Big(1-\frac{\pi^2}{6 k^2}(N_1^2+N_2^2-2)\ +O\left(\frac{1}{k^3}\right)\Big) 
\end{equation}
It is then easy to see that this expression coincides with the perturbative result \eqref{pertresult} integrated twice on the great circle. 
We have thus checked identity \eqref{identity} at perturbative level. 

Just to complete the picture, the central charge in \eqref{centralcharge3} indeed satisfies the identity
\begin{equation}
c_T=-\frac{64}{\pi^2} \, \frac{\partial^2}{\partial m^2_{-}} \log{Z[S^3, m_\pm]}\bigg|_{m_\pm=0}
\end{equation}
in agreement with the general finding of \cite{Closset:2012vg}.

\subsection{A prediction for the four-point function at two loops}

From the general structure of the partition function in \eqref{Partitionmatrix} it is easy to see that all the odd-order mass derivatives evaluated at $m_\pm=0$ vanish identically due to symmetry reasons. Therefore, odd topological correlators should vanish at any order in loops, in agreement with our findings of section \ref{sect: perturbative}. Even number of derivatives are instead non-vanishing and can be used to obtain predictions for even topological correlators at weak coupling. 

Here we consider the simplest case beyond the two-point function, that is the connected four-point function. Generalizing the prescription in \eqref{identity} for the two-point function in an obvious way, we can write
\begin{equation}
\langle\mathcal{O}(\tau_1)\mathcal{O}(\tau_2)\mathcal{O}(\tau_3)\mathcal{O}(0) \rangle =\frac{1}{\pi^4(2\pi)^4} \; \frac{\partial^4\log Z}{\partial m^4_{-}}\bigg|_{m_{-}=0}=\frac{1}{\pi^4(2\pi)^4} \; \left(\frac{\mathcal Z''''}{\mathcal Z}-3\left(\frac{\mathcal Z''}{\mathcal Z}\right)^2  \right) 
\end{equation}
where we have used $Z'=0$ and normalized everything by $Z_0$, eq. \eqref{Z0}. The second term can be easily recognized to be three times the square of the two-point function \eqref{twoderivative}, thus this expression computes correctly the connected correlation function. 

Evaluating explicitly ${\cal Z}''''$ at order $g_s^4$ and using \eqref{ZZsec}, we obtain a two-loop prediction for the four-point topological correlator 
\begin{equation}
\langle \mathcal{O}(\tau_1)\mathcal{O}(\tau_2)\mathcal{O}(\tau_3)\mathcal{O}(0) \rangle=2\frac{N_1N_2}{(4\pi)^4}-\frac{N_1N_2(N_1^2+N_2^2-2)}{192\pi^2 \ k^2}+O\left(\frac{1}{k^3}\right)
\end{equation}
We note that up to one loop it agrees with our perturbative result \eqref{4pt}, whereas the $\frac{1}{k^2}$ term is a new non-trivial result which should be checked against a genuine two--loop calculation. 
 
\section{Conclusions and future directions}\label{conclusions}

In this paper we have investigated the one-dimensional topological sector of ${\cal N} = 6$ ABJ(M) theory, taking a slightly different point of view with respect to previous investigations \cite{Binder:2019mpb,Binder:2020ckj}. We started directly from the superconformal algebra 
$\mathfrak{su}(1,1|3)\oplus\mathfrak{u}(1)_b$, the symmetry of the straight line, and we obtained the relevant cohomology working directly in the $\mathfrak{su}(1,1|3)$ formalism. The topological operators have been identified with some superconformal primaries and constructed explicitly as composite operators of the fundamental matter fields of the theory. Then we turned our attention to the correlation functions of dimension-one topological operators, which have the nice property inherited from the full $\mathfrak{osp}(6|4)$ algebra, of being related to the correlators of the stress-energy tensor \cite{Binder:2019mpb}. Moreover they are simple enough to be studied perturbatively at loop level, using conventional Feynman diagrams. We have computed the two-point correlation function at two loops and found perfect agreement with the second derivative of the mass-deformed partition function of ABJ(M) theory, evaluated at weak coupling directly from its Matrix Model representation. Our result supports the proposal in \cite{Binder:2019mpb} that the mass-deformed partition function is a sort of generating functional for the (integrated) correlation functions. As a by-product we obtained the explicit expression for the central charge $c_T$ at the second perturbative order, for generic $N_1, N_2$. 

Three- and four-point functions at one loop have also been proved to be consistent with the Matrix Model results. Pushing their computation at two-loop would certainly enforce our confidence with the proposed relation. 

For ${\cal N} = 4$ SCFTs without Chern-Simons terms the correct framework to link the topological sector and the mass-deformed theory is to perform localization with the supercharges used in \cite{Dedushenko:2016jxl}. This procedure directly generates a topological one-dimensional quantum 
mechanics governing the topological correlation functions of the full theory. In the ${\cal N} = 6$ case localization has been performed so far only with the usual ${\cal N} = 2$  KYW supercharges and it is an open problem to extend the approach of \cite{Dedushenko:2016jxl} to the ABJ(M) theory\footnote{A similar situation is suffered by the latitude Wilson loop in ABJM \cite{Cardinali:2012ru}: a conjectured matrix-model has been proposed in \cite{Bianchi:2018bke} to compute its expectation value but no first principle derivation has been found so far. See anyway \cite{Griguolo:2021rke} for progress in this direction.}. Nevertheless the emerging picture seems to suggest the possibility that a topological quantum mechanics could emerge from some localization procedure, effectively describing the full topological sector of ${\cal N} = 6$ super Chern-Simons theories, including operators of arbitrary dimensions and possibly the monopole sector \cite{Dedushenko:2017avn}.

A natural generalization of the present investigation would concern the construction of topological operators inserted into the 1/2 BPS Wilson line. Defect conformal field theories supported on the 1/2 BPS Wilson line have been studied in four-dimensional ${\cal N}=4$ SYM \cite{Giombi:2017cqn,Liendo:2018ukf} and its topological sector has been extensively studied in a series of papers \cite{Giombi:2018qox,Giombi:2018hsx,Giombi:2020amn}. The defect conformal field theory related to the 1/2 BPS Wilson line in ABJ(M) theory has been examined in \cite{Bianchi:2020hsz}, where it has been shown that, at variance with the four-dimensional case,  the displacement supermultiplet does not admit a topological sector. Because the relevant symmetry is exactly $\mathfrak{su}(1,1|3)\oplus\mathfrak{u}(1)_b$ we expect that an explicit representation of the topological operators can be constructed, although in terms of supermatrices, as done in \cite{Bianchi:2020hsz} for the displacement supermultiplet. Work in this direction is in progress \cite{forthcoming}. 

Another interesting perspective would be to apply conformal bootstrap techniques in this context. In the ${\cal N}=4$ case the OPE data in
the relevant topological quantum mechanics can be obtained or constrained imposing the associativity and unitarity of the operator algebra 
\cite{Beem:2016cbd,Chang:2019dzt}. This procedure is dubbed mini-bootstrap (or micro-bootstrap in four-dimensions \cite{Liendo:2016ymz}) because it concerns a closed subsystem of the full bootstrap equations. The generalization to the ${\cal N}\geq4$ case could give further hints on the structure of the topological quantum mechanics and might produce new solutions corresponding to presently unknown sectors.

\vskip 50pt
\noindent
{\bf Acknowledgments}

\vskip 10pt
\noindent
We thank Lorenzo Bianchi,  Carlo Meneghelli and Itamar Yaakov for stimulating discussions. We are grateful to Marco Bianchi for his collaboration on section \ref{sect: perturbative}. This work has been supported in part by Italian Ministero dell’Istruzione, Universit\`a e Ricerca (MIUR), and Istituto Nazionale di Fisica Nucleare (INFN) through the “Gauge and String Theory” (GAST) and “Gauge Theories, Strings, Supergravity” (GSS) research projects.

\newpage
\appendix
\section{ABJ(M) action and Feynman rules}\label{ABJ(M)}

We work in euclidean space with coordinates $x^\mu=(x^1,x^2,x^3)$ and metric $\delta_{\mu\nu}$. 
Gamma matrices satisfying the usual Clifford algebra $\{\gamma^\mu,\gamma^\nu\}=2\delta^{\mu\nu}\mathbb 1$, are chosen to be the Pauli matrices 
\beq
(\gamma^\mu)_\alpha^{\; \beta} \equiv (\sigma^\mu)_\alpha^{\; \beta} \qquad \quad \mu = 1,2,3
\eeq
Standard relations which are useful for perturbative calculations  are
\bea
\gamma^\mu\gamma^\nu &=& \delta^{\mu\nu}+i\varepsilon^{\mu\nu\rho}\gamma_\rho \\
\g^\mu \g^\nu \g^\rho &=& \delta^{\mu\nu} \g^\rho - \delta^{\mu\rho} \g^\nu + \delta^{\nu\rho} \g^\mu + i \varepsilon^{\mu\nu\rho}
\eea 
Moreover, we define $\gamma^{\mu\nu}\equiv\frac{1}{2}[\gamma^\mu,\gamma^\nu]$.

Spinor indices are raised and lowered according to the following rules
\[
\psi^\alpha=\varepsilon^{\alpha\beta}\psi_\beta,\qquad \psi_\alpha=\varepsilon_{\alpha\beta}\psi^\beta
\]
with $\varepsilon^{12}=-\varepsilon_{12}=1$. Consequently, we define $(\g^\mu)_{\a \b } \equiv \varepsilon_{\b \g} (\g^\mu)_\a^{\; \g} = (-\sigma^3, i {\rm I}, \sigma^1)$ and $(\g^\mu)^{\a \b } \equiv \varepsilon^{\a \g} (\g^\mu)_\g^{\; \b} = (\sigma^3, i {\rm I}, -\sigma^1)$. They satisfy $(\g^\mu)_{\a \b } = (\g^\mu)_{\b \a }$ and $(\g^\mu)^{\a \b } = (\g^\mu)^{\b \a }$.

\vskip 10pt

The $U(N_1)_k \times U(N_2)_{-k}$ ABJ(M) theory contains two gauge fields $(A_\mu)_i^j$, $(\hat A_\mu)_{\hat i}^{\hat j}$ belonging to the adjoint representation of $U(N_1)$ and $U(N_2)$ respectively, minimally coupled to matter realised in terms of four multiplets $(C_I, \bar{\psi}^I)_{I=1, \dots , 4}$ in the $(N_1, \bar{N}_2)$ representation of the gauge group and their conjugates $(\bar{C}^I, \psi_I)_{I=1, \dots , 4}$ in the 
$(\bar{N}_1, N_2)$. 

We use conventions in \cite{Bianchi:2018bke} with a convenient rescaling of the gauge fields and the corresponding ghosts,  \(A\rightarrow \frac{1}{\sqrt{k}}A,\;\hat{A}\rightarrow\frac{1}{\sqrt{k}}\hat{A}\),  \(c\rightarrow \frac{1}{\sqrt{k}}c,\;\hat{c}\rightarrow \frac{1}{\sqrt{k}}\hat{c}\).  Defining covariant derivatives as 
\begin{equation}
\begin{aligned}
&D_\mu C_I =\de_\mu C_I +\frac{i}{\sqrt{k}}  A_\mu C_I- \frac{i}{\sqrt{k}}  C_I \hat{A}_\mu ,\qquad D_\mu \bar{C}^I =\de_\mu \bar{C}^I +\frac{i}{\sqrt{k}}  \hat{A}_\mu \bar{C}^I-\frac{i}{\sqrt{k}} \bar{C}^I A_\mu \\
&D_\mu \bar{\psi}^I =\de_\mu \bar{\psi}^I +\frac{i}{\sqrt{k}}  A_\mu \bar{\psi}^I-\frac{i}{\sqrt{k}} \bar{\psi}^I \hat{A}_\mu ,\qquad D_\mu \psi_I =\de_\mu \psi_I +\frac{i}{\sqrt{k}}  \hat{A}_\mu \psi_I-\frac{i}{\sqrt{k}} \psi_I A_\mu 
\end{aligned}
\label{covd}
\end{equation}
the Euclidean gauge-fixed action is then given by 
\begin{equation}\label{action}
S=S_{\rm CS}+S_{\rm gf}+S_{\rm mat} + S_{\rm pot}
\end{equation}
where
\bea
S_{\rm CS} &=&  -\frac{i}{4\pi}\int d^3x \ \varepsilon^{\mu\nu\rho}\left[
\Tr\left(A_\mu\de_\nu A_\rho+\frac{2i}{3\sqrt{k}}A_\mu A_\nu A_\rho\right)-\Tr\left(\hat{A}_\mu\de_\nu\hat{A}_\rho+\frac{2i}{3\sqrt{k}}\hat{A}_\mu\hat{A}_\nu\hat{A}_\rho\right)\right] \non \\
&& \label{CS} \\
S_{\rm gf} &=& \frac{1}{4\pi}\int d^3x \ \Tr\left[\frac{1}{\xi}\left(\de_\mu A^\mu\right)^2+\de_\mu\bar{c}D^\mu c-\frac{1}{\xi}\left(\de_\mu\hat{A}^\mu\right)^2-\de_\mu\bar{\hat{c}}D^\mu\hat{c}\right] \\ 
S_{\rm mat} &=& \int d^3x \ \Tr\left[D_\mu C_I D^\mu\bar{C}^I - i\bar{\psi}^I\gamma^\mu D_\mu\psi_I\right] \\
&= & \int d^3x \ \Tr\left[\de_\mu C_I\de^\mu \bar{C}^I - i\bar{\psi}^I\gamma^\mu\de_\mu\psi_I+\frac{1}{\sqrt{k}}\left(\bar{\psi}^I\gamma^\mu\hat{A}_\mu\psi_I 
- \bar{\psi}^I\gamma^\mu\psi_I A_\mu\right) \right. \notag \\
&&\left.\qquad\qquad+\frac{i}{\sqrt{k}}\left(A_\mu C_I\de^\mu\bar{C}^I-C_I\hat{A}_\mu\de^\mu \bar{C}^I-\de_\mu C_I\bar{C}^I A^\mu+\de_\mu C_I\hat{A}^\mu\bar{C}^I\right) \right. \notag \\
&&\left.\qquad\qquad+\frac{1}{k}\left(A_\mu C_I\bar{C}^I A^\mu-A_\mu C_I \hat{A}^\mu\bar{C}^I-C_I \hat{A}_\mu\bar{C}^I A^\mu+C_I\hat{A}_\mu\hat{A}^\mu\bar{C}^I\right)\right] \label{Smat}
\eea
\bea 
&& \hspace{-1.8cm} S_{\rm pot} \equiv  S_{\rm 6pt} + S_{\rm 4pt} \non \\
&&  \hspace{-1.8cm} \qquad  = -\frac{4\pi^2}{3 k^2} \int d^3x \, \Tr \Big[ C_I \bar{C}^I C_J \bar{C}^J C_K \bar{C}^K + \bar{C}^I C_I \bar{C}^J C_J \bar{C}^K C_K
\non \\
&&  \hspace{-1.8cm} ~  \qquad \qquad \qquad \qquad  \quad + 4 \, C_I \bar{C}^J C_K \bar{C}^I C_J \bar{C}^K - 6 \, C_I \bar{C}^J C_J \bar{C}^I C_K \bar{C}^K \Big] 
\label {S6pt}\\
&&  \hspace{-0.3cm} -\frac{2\pi i}{k} \int d^3x \, \Tr \Big[ \bar{C}^I C_I \Psi_J \bar{\Psi}^J - C_I \bar{C}^I \bar{\Psi}^J \Psi_J
+2 \, C_I \bar{C}^J \bar{\Psi}^I \Psi_J 
\non \\
&&  \qquad \qquad \quad    - 2 \, \bar{C}^I C_J \Psi_I \bar{\Psi}^J - \epsilon_{IJKL} \bar{C}^I\bar{\Psi}^J \bar{C}^K \bar{\Psi}^L + \epsilon^{IJKL} C_I \Psi_J C_K \Psi_L \Big]
\label{S4pt}
\eea
where $\epsilon_{1234}=\epsilon^{1234} =1$ and for the group generators we use the following relations
\beq
\Tr (T^A T^B) = \delta^{AB} \; , \qquad [ T^A , T^B ] = i f^{AB}_{\; \; \; \; \, C} \, T^C
\eeq

\vskip 15pt

The corresponding propagators at tree and loop orders, as needed for the two-loop calculations, are:
\begin{itemize}
	\item Scalar propagator
	\begin{align}
	\langle {(C_I)_i}^{\hat{j}} (x)\ {(\bar C^J)_{\hat k}}^l (y)  \rangle^{(0)}&=\delta^J_I\delta^l_i\delta^{\hat j}_{\hat k} \ \frac{\Gamma(\frac{1}{2}-\epsilon)}{{4\pi}^{\frac{3}{2}-\epsilon}}\frac{1}{{|x-y|}^{1-2\epsilon}} \label{scalartree}\\
	\langle {(C_I)_i}^{\hat{j}} (x)\ {(\bar C^J)_{\hat k}}^l (y)  \rangle^{(1)}&=0 
	\end{align}  
	\item Vector propagators in Landau gauge
\bea\label{prop:vector}
	&& \langle {(A_\mu)_i}^j (x)\ {(A_\nu)_k}^l (y)  \rangle^{(0)}=  \delta^l_i\delta^j_k \   i  \, \frac{\Gamma(\frac{3}{2}-\epsilon)}{{\pi}^{\frac{1}{2}-\epsilon}}\ \varepsilon_{\mu\nu\rho}\ \frac{(x-y)^\rho}{{|x-y|}^{3-2\epsilon}} \non \\
	&&	\langle {(\hat A_\mu)_{\hat i}}^{\hat j} (x)\ {(\hat A_\nu)_{\hat k}}^{\hat l} (y)  \rangle^{(0)}=  - \, \delta^{\hat l}_{\hat i}\delta^{\hat j}_{\hat k} \   i  \, \frac{\Gamma(\frac{3}{2}-\epsilon)}{{\pi}^{\frac{1}{2}-\epsilon}}\ \varepsilon_{\mu\nu\rho}\ \frac{(x-y)^\rho}{{|x-y|}^{3-2\epsilon}}
\eea
\bea
&& \hspace{-1cm} \langle {(A_\mu)_i}^j (x)\ {(A_\nu)_k}^l (y)  \rangle^{(1)} = \delta^l_i\delta^j_k \    \frac{N_2}{k} \, \frac{\Gamma^2(\frac12-\e)}{\pi^{1 -2\e}} 
\left( \frac{\delta_{\mu\nu}}{ |x- y|^{2-4\e}} - \pa_\mu \pa_\nu \frac{|x-y|^{4\e}}{4\e(1+2\e)} \right)  
\non \\
&& \hspace{-1cm} \langle {(\hat A_\mu)_{\hat i}}^{\hat j} (x)\ {(\hat A_\nu)_{\hat k}}^{\hat l} (y)  \rangle^{(1)} = \delta^l_i\delta^j_k \  \frac{N_1}{k} \, \frac{\Gamma^2(\frac12-\e)}{\pi^{1 -2\e}} 
\left( \frac{\delta_{\mu\nu}}{ |x- y|^{2-4\e}} - \pa_\mu \pa_\nu \frac{|x-y|^{4\e}}{4\e(1+2\e)} \right) \non \\
\eea	
	\item Fermion propagator
	\begin{equation}
	\langle {(\psi_{\alpha I})_{\hat i}}^j (x)\ {(\bar \psi^{J\beta})^{\hat l}}_k (y)  \rangle^{(0)}= \delta^J_I\delta^{\hat l}_{\hat i}\delta^j_k \ i \, \frac{\Gamma(\frac{3}{2}-\epsilon)}{{2\pi}^{\frac{3}{2}-\epsilon}} \ {(\gamma^\mu)_\alpha}^\beta\ \frac{(x-y)_\mu}{{|x-y|}^{3-2\epsilon}}
	\end{equation}
\bea
\label{1fermion}
&&\langle {(\psi_{\alpha I})_{\hat i}}^j (x)\ {(\bar \psi^{J\beta})^{\hat l}}_k (y)  \rangle^{(1)}=  -  \delta_I^J \delta_{\hat{i}}^{\hat{l}} \delta_{k}^{j} 
\,  \, \delta_\a^{\; \, \b} \,  \left( \frac{N_1 - N_2}{k} \right) \, i \, \frac{\Gamma^2(\frac12 - \e)}{8 \pi^{2-2\e}} \, \frac{1}{|x-y|^{2 - 4\e}}  
\non \\
\eea 	
We note that in the ABJ(M) limit, $N_1 = N_2$, the one-loop correction to the fermionic propagator vanishes. 
\end{itemize}

The vertices entering the perturbative calculations of section [\ref{sect: perturbative}] can be easily read from terms \eqref{CS}, \eqref{Smat} and \eqref{S4pt} of the action.

\vskip40pt

\section{Euclidean $\mathfrak{osp}(6|4)$ superalgebra}\label{osp64} 
In Euclidean signature the generators of the bosonic conformal algebra contained in the $\mathfrak{osp}(6|4)$ superalgebra satisfy the following commutation rules
\begin{equation}
\begin{aligned}\label{balgebra}
[M^{\mu\nu}, M^{\rho\sigma}]&=\delta^{\sigma\mu}M^{\nu\rho}-\delta^{\sigma\nu}M^{\mu\rho}+\delta^{\rho\nu}M^{\mu\sigma}-\delta^{\rho\mu}M^{\nu\sigma}\ &[P^\mu,K^\nu]&=2(\delta^{\mu\nu}D+M^{\mu\nu}) \\
[P^\mu,M^{\nu\rho}]&=\delta^{\mu\nu}P^{\rho}-\delta^{\mu\rho}P^\nu &[K^\mu,M^{\nu\rho}]&=\delta^{\mu\nu}K^{\rho}-\delta^{\mu\rho}K^\nu\\
[D,P^\mu]&=P^\mu &[D,K^\mu]&=-K^\mu
\end{aligned}
\end{equation}
The $\mathfrak{su}(4)\simeq\mathfrak{so}(6)$ R-symmetry generators ${J_I}^J$, with $I,J=1,\dots,4$, are traceless matrices that satisfy the relation
\begin{equation}\label{Ralgebra}
[{J_I}^J, {J_K}^L]=\delta^L_I {J_K}^J-\delta^J_K {J_I}^L
\end{equation}
The fermionic generators $Q^{IJ}_\alpha$, $S^{IJ}_\alpha$ satisfy the following anticommutation rules
\begin{equation}\label{QSalgebra}
\begin{aligned}
\{Q^{IJ}_\alpha, Q^{KL,\beta}\}&= \varepsilon^{IJKL}{(\gamma^\mu)_{\alpha}}^\beta P_\mu\qquad\qquad \{S_\alpha^{IJ}, S^{\beta KL}\}= \varepsilon^{IJKL}{(\gamma^\mu)_{\alpha}}^\beta K_\mu\\
\{Q^{IJ}_\alpha, S^{\beta KL}\}&= \varepsilon^{IJKL}\left(\frac12 {(\gamma^{\mu\nu})_\alpha}^\beta M_{\mu\nu}+\delta^\beta_\alpha D\right)+\delta^\beta_\alpha \varepsilon^{KLMN}(\delta^J_M {J_N}^I-\delta^I_M {J_N}^J)   
\end{aligned}
\end{equation}
and similarly for $\bar Q_{\alpha IJ}=\frac{1}{2}\varepsilon_{IJKL}Q^{KL}_\alpha$ and $\bar S_{\alpha IJ}=\frac{1}{2}\varepsilon_{IJKL}S^{KL}_\alpha$. 

The full $\mathfrak{osp}(6|4)$ superalgebra is obtained by completing the picture with the mixed commutators
\begin{equation}
\begin{aligned}
[K^\mu,Q_{\alpha}^{IJ}]&={(\gamma^\mu)_\alpha}^\beta S^{ IJ}_\beta &[P^\mu,S_{\alpha}^{IJ}]&={(\gamma^\mu)_\alpha}^\beta Q^{ IJ}_\beta\\
[M^{\mu\nu}, Q_{\alpha}^{IJ}]&=-\frac{1}{2}{(\gamma^{\mu\nu})_\alpha}^\beta Q_{\beta}^{IJ} &
[M^{\mu\nu}, S_{\alpha}^{IJ}]&=-\frac{1}{2}{(\gamma^{\mu\nu})_\alpha}^\beta S_{\beta}^{IJ}\\
[D, Q_{\alpha}^{IJ}]&=\frac{1}{2}Q_{\alpha}^{IJ} &[D, S^{\alpha IJ}]&=-\frac{1}{2}S^{\alpha IJ} \\
[{J_I}^J, Q^{KL}_\alpha]&=\delta^K_I Q^{JL}_\alpha+\delta^L_I Q^{KJ}_\alpha-\frac{1}{2}\delta^J_IQ^{KL}_\alpha &[{J_I}^J, S^{\alpha KL}]&=\delta^K_I S^{\alpha JL}+\delta^L_I S^{\alpha KJ}-\frac{1}{2}\delta^J_I S^{\alpha KL}
\end{aligned}
\end{equation}
The bosonic generators in \eqref{balgebra}, \eqref{Ralgebra} are taken to satisfy the following conjugation rules
\beq
(P^\mu)^\dagger = - K^\mu \, \qquad (K^\mu)^\dagger = - P^\mu \, \qquad D^\dagger = D \, \qquad (M^{\mu\nu})^\dagger = - M^{\mu\nu} \, \qquad ({J_K}^L)^\dagger={J_L}^K
\eeq
whereas the fermionic ones are subject to the following hermicity conditions
\begin{equation}\label{hermit}
(Q^{IJ}_\alpha)^\dagger=\frac{1}{2}\varepsilon_{IJKL}\ S^{KL\alpha} = \bar{S}_{IJ}^\a \qquad \qquad 
(S^{IJ}_\alpha)^\dagger=\frac{1}{2}\varepsilon_{IJKL}\ Q^{KL \alpha} = \bar{Q}_{IJ}^\a
\end{equation}

\noindent
The action of the $\mathfrak{su}(4)$ R-symmetry generators on fields $\Phi_I$ ($\bar{\Phi}^I$) in the (anti-)fundamental representation reads
\begin{equation}\label{fund4}
[{J_I}^J, \Phi_K] =\frac{1}{4}\delta^J_I \Phi_K-\delta^J_K \Phi_I \qquad \qquad 
[{J_I}^J, \bar{\Phi}^K]=\delta^K_I \bar{\Phi}^J - \frac{1}{4}\delta^J_I \bar{\Phi}^K
\end{equation}
The full analysis of the relevant multiplets of $\mathfrak{osp}(6|4)$ is discussed in \cite{Liendo:2015cgi}.

\vskip40pt

\section{The $\mathfrak{su}(1,1|3)$ superalgebra}\label{sect: su(1,1|3)}

In this appendix we describe the immersion of the $\mathfrak{su}(1,1|3)$ superalgebra inside $\mathfrak{osp}(6|4)$ and the classification of its irreducible representations. 

\subsection{The superalgebra}

The maximal bosonic subalgebra of $\mathfrak{su}(1,1|3)$ is $\mathfrak{sl}(2) \oplus \mathfrak{su}(3) \oplus \mathfrak{u}(1)$, where $\mathfrak{sl}(2)\simeq\mathfrak{su}(1,1)$ is the euclidean conformal algebra in one dimension and $\mathfrak{su}(3) \oplus \mathfrak{u}(1)$ is the R-symmetry algebra. 

The  $\mathfrak{su}(1,1)$ algebra is generated by $\{ P \equiv iP_3,  K\equiv iK_3, D \}$ satisfying the following commutation relations
\begin{equation}\label{su1,1}
[D,P]=P \ \qquad [D,K]=-K \ \qquad [P,K]=-2D    
\end{equation}
The $\mathfrak{su}(3)$ R-symmetry subalgebra is generated by traceless operators ${R_a}^b$, whose explicit form reads
\begin{equation}\label{Rgenerators}
{R_a}^b=
\renewcommand\arraystretch{1.2}\begin{pmatrix}
{J_2}^2+\frac{1}{3}{J_1}^1 & {J_2}^3 & {J_2}^4 \\
{J_3}^2 & {J_3}^3+\frac{1}{3}{J_1}^1 &  {J_3}^4 \\
{J_4}^2 &  {J_4}^3 &-{J_3}^3-{J_2}^2-\frac{2}{3}{J_1}^1  
\end{pmatrix}
\end{equation}
These generators satisfy the algebraic relation
\begin{equation}\label{su3}
[{R_a}^b, {R_c}^d]=\delta_a^d {R_c}^b- \delta_c^b {R_a}^d     
\end{equation}
The spectrum of bosonic generators of $\mathfrak{su}(1,1|3)$ is completed by a residual $\mathfrak{u}(1)$ generator $M$, defined as
\begin{equation}\label{M}
M\equiv 3i M_{12}-2{J_1}^1    
\end{equation}

We now move to the fermionic sector of the superalgebra. Since we have placed the line along the $x^3$-direction, the fermionic generators of the one-dimensional superconformal algebra are identified with the following supercharges
\begin{equation}
Q_1^{12}, Q_1^{13}, Q^{14}_1, Q_2^{23}, Q^{24}_2, Q^{34}_2\qquad\text{and}\qquad S_1^{12}, S_1^{13}, S^{14}_1, S_2^{23}, S^{24}_2, S^{34}_2
\end{equation}

It is useful to rewrite these generators in a more compact way, through the following definitions
\begin{equation}
\begin{aligned}\label{supercharges}
Q^{k-1}&\equiv Q^{1 k}_1 \qquad 
&\bar Q_{k-1}&\equiv \frac{i}{2} \, \epsilon_{klm} \, Q^{lm}_2  \\
S^{k-1}&\equiv i \,  S^{1k}_1 &
\bar S_{k-1} &\equiv \frac{1}{2} \, \epsilon_{klm} \, S^{lm}_2 \qquad \quad k,l,m =2,3,4
\end{aligned}
\end{equation}
and make the shift $Q^{k-1} \to Q^{a}$,  $\bar Q_{k-1} \to \bar Q_{a}$ with $a=1,2,3$, and similarly for the superconformal charges. 

This set of generators inherits the following hermicity conditions
\begin{equation}
\begin{aligned}
 (Q^a)^\dagger&=\bar S_a\qquad \ &(\bar Q_a)^\dagger&=S^a\\
 (S^a)^\dagger&=\bar Q_a &(\bar S_a)^\dagger&=Q^a \qquad \quad a=1,2,3
\end{aligned}
\end{equation}
and the following anti-commutation relations 
\begin{equation}\label{anticomm}
\begin{aligned}
\{Q^a, \bar Q_b\}&= \delta^a_b \, P\qquad \quad &\{S^a, \bar S_b\} &=\delta^a_b \, K \\
\{Q^a, \bar S_b\}&= \delta^a_b\bigg(D+\frac{1}{3}M\bigg)- {R_b}^a &
\{\bar Q_a, S^b\}&= \delta_a^b\bigg(D-\frac{1}{3}M\bigg)+ {R_a}^b
\end{aligned}
\end{equation}
together with the mixed commutation rules
\begin{equation}\label{comm}
\begin{aligned} 
[D,Q^a]&=\frac{1}{2}Q^a\qquad &[K,Q^a]&=S^a\qquad &[{R_a}^b, Q^c]&=\delta^c_a Q^b-\frac{1}{3}\delta_a^b Q^c\qquad & [M, Q^a]&=\frac{1}{2}Q^a\\ 
[D,\bar Q_a]&=\frac{1}{2}\bar Q_a \qquad
 &[K,\bar Q_a]&=\bar S_a &[{R_a}^b, \bar Q_c]&=-\delta^b_c \bar Q_a+\frac{1}{3}\delta_a^b \bar Q_c &[M, \bar Q_a]&=-\frac{1}{2}\bar Q_a    \\ 
[D, S^a]&=-\frac{1}{2}S^a\qquad 
 &[P, S^a]&=-Q^a &[{R_a}^b, S^c]&=\delta^c_a S^b-\frac{1}{3}\delta_a^b S^c &[M,S^a]&=\frac{1}{2}S^a  \\
[D, \bar S_a]&=-\frac{1}{2}\bar S_a &[P, \bar 
S_a]&=-\bar Q_a &[{R_a}^b, \bar S_c]&=-\delta^b_c \bar S_b+\frac{1}{3}\delta_a^b \bar S_c & [M, \bar S_a]&=-\frac{1}{2}\bar S_a \\
& \ 
\end{aligned}
\end{equation}

From eq. \eqref{fund4} and definitions \eqref{Rgenerators} it follows that the action of the $SU(3)$ R-symmetry generators on fields in the (anti-)fundamental representation is
\begin{equation}\label{fund3}
[{R_a}^b, \Phi_c]=\frac{1}{3}\delta^b_a \Phi_c-\delta^b_c \Phi_a \qquad  [{R_a}^b, \bar{\Phi}^c]=\delta^c_a \bar{\Phi}^b - \frac{1}{3}\delta^b_a \bar{\Phi}^c
\end{equation}

\vskip 10pt

\subsection{Irreducible representations}\label{irrepline}
In this appendix, we shall briefly review the classification of the multiplet of  $\mathfrak{su}(1,1|3)$ presented in \cite{Bianchi:2017ozk}.
We shall classify the states in terms of the  four Dynkin labels $[\Delta, m, j_1, j_2]$ associated to the bosonic subalgebra $\mathfrak{su}(1,1)\oplus  \mathfrak{su}(3) \oplus \mathfrak{u}(1)$. Here $\Delta$ stands for  the conformal weight, $m$ for the $\mathfrak{u}(1)$ charge and $(j_1, j_2)$ are the eigenvalues corresponding to the two $\mathfrak{su}(3)$ Cartan generators $J_1$ and $J_2$. We choose
\begin{equation}\label{su(3)cartan}
\begin{aligned}
& J_1\equiv \frac{{R_2}^2-{R_1}^1}{2} =  - \frac{2{R_1}^1 + {R_3}^3}{2} \\
&J_2\equiv \frac{{R_3}^3-{R_2}^2}{2} =  \frac{{R_1}^1 + 2{R_3}^3}{2} 
\end{aligned}
\end{equation}
where we have exploited the traceless property ${R_a}^a=0$ to remove the dependence on ${R_2}^2$. 
The commutations rules \eqref{su3} implies that we can associate an $\mathfrak{sl}(2)$ subalgebra with each Cartan generator.
 In fact, the two sets of operators
\begin{equation}
\begin{aligned}
\{ {R_2}^1, {R_1}^2, J_1\} \equiv \{ E^{-}_1, E^+_1, J_1 \}   \qquad , \qquad 
\{ {R_3}^2, {R_2}^3, J_2\} \equiv \{ E^{-}_2, E^+_2, J_2 \}  
\end{aligned}
\end{equation}
satisfy the following algebraic relations 
\begin{equation}
[E_i^+,E_i^-]=2J_i\qquad  [J_i,E_i^{\pm}]=\pm E_i^{\pm}  \qquad \quad i = 1,2
\end{equation}
and  define the raising and lowering operators used to construct the representation of $\mathfrak{su}(3)$. In the main text, we have  chosen a different  $\mathfrak{sl}(2)$ to define the twisted algebra. We have preferred to use the one generated by $\{ {R_3}^1, {R_1}^3, -J_1-J_2\}$, which treats the two  Dynkin labels $(j_1,j_2)$ symmetrically.\\
Moreover,  the supercharges with this choice of basis possess well-defined Dynkin labels, whose values are displayed in Table \ref{table1}.
\begin{table}[h!]
\begin{center}
\begin{tabular}{ |c|c| } 
 \hline
 {\rm Generators} & $[\Delta, m, j_1, j_2]$  \\ 
 \hline\hline
 $Q^1$ $\; \bar Q_1$ & $\left[\frac{1}{2},\frac{1}{2},-1,0\right]$ \quad $\left[\frac{1}{2},-\frac{1}{2},1,0\right]$ \\  
 $Q^2$ $\; \bar Q_2$ & $\left[ \frac{1}{2},\frac{1}{2},1,-1\right] $  \quad $\left[\frac{1}{2},-\frac{1}{2},-1,1\right]$ \\ 
 $Q^3$ $\; \bar Q_3$ & $\left[\frac{1}{2},\frac{1}{2},0,1\right]$  \quad $\left[\frac{1}{2},-\frac{1}{2},0,-1\right]$ \\
\hline\hline
$S^1$ $\; \bar S_1$ &  $\left[-\frac{1}{2},\frac{1}{2},-1,0\right]$ \quad  $\left[-\frac{1}{2},-\frac{1}{2},1,0\right]$ \\
$S^2$ $\; \bar S_2$ & $\left[- \frac{1}{2},\frac{1}{2},1,-1\right]$ \quad $\left[-\frac{1}{2},-\frac{1}{2},-1,1\right]$ \\
$S^3$ $\; \bar S_3$ & $\left[-\frac{1}{2},\frac{1}{2},0,1\right]$ \quad $\left[-\frac{1}{2},-\frac{1}{2},0,-1\right]$ \\
 \hline
\end{tabular}
\caption{Table of Dynkin labels of fermionic generators. For a generic element $v_\mu$ transforming in a weight-$\mu$ representation, the Dynkin label corresponding to a generator  $H_{i}$ of the Cartan subalgebra is defined as $j_i (v_\mu) \equiv 2[H_{i}, v_\mu]$. }
\label{table1}
\end{center}
\end{table}
\vskip-0.7cm
\noindent
When  localized on the line, the ABJ(M) fundamental fields  also have definite  quantum numbers with respect to $\mathfrak{su}(1,1)\oplus  \mathfrak{su}(3) \oplus \mathfrak{u}(1)$. Their values are listed in Table \ref{table2} for the scalar fields and in Table \ref{table3} for the fermionic ones.

\begin{table}[h!]
\begin{center}
\begin{tabular}{ |c|c| } 
 \hline
 {\rm Scalar fields} & $[\Delta, m, j_1, j_2]$  \\ 
 \hline\hline
 $Z$ $, \; \bar Z$ & $\left[\frac{1}{2},\frac{3}{2},0,0\right]$ \quad $\left[\frac{1}{2},-\frac{3}{2},0,0\right]$ \\  
 $Y_1$ $,\; \bar Y^1$ & $\left[ \frac{1}{2},-\frac{1}{2},1,0\right] $  \quad $\left[\frac{1}{2},\frac{1}{2},-1,0\right]$ \\ 
 $Y_2$ $,\; \bar Y^2$ & $\left[\frac{1}{2},-\frac{1}{2},-1,1\right]$  \quad $\left[\frac{1}{2},\frac{1}{2},1,-1\right]$ \\
  $Y_3$ $,\; \bar Y^3$ & $\left[\frac{1}{2},-\frac{1}{2},0,-1\right]$  \quad $\left[\frac{1}{2},\frac{1}{2},0,1\right]$ \\
\hline
 \hline
\end{tabular}
\caption{Quantum number assignments to scalar matter fields of the ABJ(M) theory defined in eq. \eqref{su3breaking}.} 
\label{table2}
\end{center}
\end{table}
\begin{table}[h!]
\begin{center}
\begin{tabular}{ |c|c| } 
 \hline
 {\rm Fermionic fields} & $[\Delta, m, j_1, j_2]$  \\ 
 \hline\hline
 $(\psi)_1$ $, \; (\psi)_2$ & $\left[1,3,0,0\right]$ \quad $\left[1,0,0,0\right]$ \\  
 $(\bar \psi)_1$ $,\; (\bar \psi)_2$ & $\left[1,0,0,0\right]$  \quad $\left[1,-3,0,0\right]$ \\ 
 $ (\chi_1)_1$ $,\; (\chi_1)_2$ & $\left[ 1,1,1,0\right]$  \quad $\left[1,-2,1,0\right]$ \\
 $ (\bar \chi^1)_1$ $,\; (\bar \chi^1)_2$ & $\left[ 1,2,-1,0\right]$  \quad $\left[1,-1,-1,0\right]$ \\
 $ (\chi_2)_1$ $,\; (\chi_2)_2$ & $\left[1,1,-1,1\right]$  \quad $\left[1,-2,-1,1\right]$ \\
 $ (\bar \chi^2)_1$ $,\; (\bar \chi^2)_2$ & $\left[1,2,1,-1\right]$  \quad $\left[1,-1,1,-1\right]$ \\
 $ (\chi_3)_1$ $,\; (\chi_3)_2$ & $\left[1,1,0,-1\right]$  \quad $\left[1,-2,0,-1\right]$ \\
 $ (\bar \chi^3)_1$ $,\; (\bar \chi^3)_2$ & $\left[1,2,0,1\right]$  \quad $\left[1,-1,0,1\right]$ \\
\hline
 \hline
\end{tabular}
\caption{Quantum number assignments to fermionic matter fields of the ABJ(M) theory defined in eq. \eqref{su3breaking}.} 
\label{table3}
\end{center}
\end{table}
\vskip -.7cm
\noindent
Finally we do not  consider directly the  gauge fields, but their covariant derivatives. Their   Dynkin labels are given by
\begin{equation}
D\ [1,3,0,0]\qquad \bar D\ [1,-3,0,0]\qquad D_3\ [1,0,0,0]    
\end{equation}
Therefore their action on an operator that is an eigenstate $| \Delta, m, j_1, j_2 \rangle$ of the Cartan generators simply  shifts the the first two quantum numbers. 
Next we summarize the relevant superconformal multiplets constructed in \cite{Bianchi:2017ozk}.
\vskip .5cm

\noindent
\textbf{The $\mathcal A$ Multiplets}

\noindent
We start with the so-called long multiplets, denoted by $\mathcal{A}^{\Delta}_{m;j_1,j_2}$.  Their highest weight of the representations, namely the super-conformal primary (SCP), is identified by requiring that
\begin{align}
 S^a\ket{\Delta,m,j_1,j_2}^{\text{hw}}&=0 & \bar S_a\ket{\Delta,m,j_1,j_2}^{\text{hw}}&=0 & E^+_a\ket{\Delta,m,j_1,j_2}^{\text{hw}}&=0
\end{align}
Then  the entire multiplet  is built by acting with the supercharges $Q^a$ and $\bar Q_a$. 
For unitary representations, the Dynkin label of the highest weight are constrained by the following inequalities
\begin{equation}\label{Amultiplet}
\Delta\ge\begin{cases}
\frac{1}{3}(2j_2+j_1-m), \qquad m<\frac{j_2-j_1}{2}\\
\frac{1}{3}(j_2+2j_1+m), \qquad m\ge\frac{j_2-j_1}{2}
\end{cases}
\end{equation}
At the threshold of the unitary region, these  multiplets split into shorter ones because of the recombination phenomenon. For $m< \frac{j_2-j_1}{2}$ the unitarity bound is for $\Delta=\frac13(2j_2+j_1-m)$ and one can verify that
\begin{equation}
 \mathcal{A}^{-\frac13 m+\frac13 j_1+\frac23 j_2}_{m,j_1,j_2}=\mathcal{B}^{\frac16,0}_{m,j_1,j_2} \oplus \mathcal{B}^{\frac16,0}_{m+\frac12,j_1,j_2+1}
\end{equation}
Equivalently, for $m> \frac{j_2-j_1}{2}$ one has
\begin{equation}
 \mathcal{A}^{\frac13 m+\frac23 j_1+\frac13 j_2}_{m,j_1,j_2}={\mathcal{B}}^{0,\frac16}_{m,j_1,j_2} \oplus {\mathcal{B}}^{0,\frac16}_{m-\frac12,j_1+1,j_2}
\end{equation}
For the particular case $m=\frac{j_2-j_1}{2}$ we have
\begin{equation}\label{recomb}
\mathcal{A}^{\frac{j_2+j_1}{2}}_{\frac{j_2-j_1}{2};j_1,j_2}=\quad\mathcal{B}^{\frac{1}{6},\frac{1}{6}}_{\frac{j_2-j_1}{2};j_1,j_2}\oplus \mathcal{B}^{\frac{1}{6},\frac{1}{6}}_{\frac{j_2-j_1}{2}+\frac{1}{2};j_1,j_2+1}\oplus \mathcal{B}^{\frac{1}{6},\frac{1}{6}}_{\frac{j_2-j_1}{2}-\frac{1}{2};j_1+1,j_2+1}\oplus \mathcal{B}^{\frac{1}{6},\frac{1}{6}}_{\frac{j_2-j_1}{2};j_1+1,j_2+1}.
\end{equation}
Above the symbols  $\mathcal{B}_{m;j_1,j_2}^{\frac{1}{N},\frac{1}{M}}$  stand for a type of short multiplets (see below). The two superscripts denote respectively the fraction of $Q$ and  $\bar Q$  charges with respect to the total number of charges ($Q+\bar Q$), which annihilates the super-conformal primary.
 
\vskip .5cm
\noindent
 \textbf{The $\mathcal B$ Multiplets}\\
Let us now have a closer look to short multiplets. They are obtained by imposing that the highest weight is   annihilated by some of the $Q$ and  $\bar Q$  charges ({\it shortening condition}). First we consider the case
\begin{equation}\label{shortening}
Q^a\ket{\Delta, m, j_1, j_2}^{\text{hw}}=0    
\end{equation}
from which we get three possible short supermultiplets
\begin{align}\label{shortening}
a&=3\qquad &\Delta&=\frac{1}{3}(j_1+2j_2-m)\qquad &\ &\mathcal{B}_{m;j_1,j_2}^{\frac{1}{6},0} \\
a&=3,2   &\Delta&=\frac{1}{3}(j_1-m),\quad j_2=0 &\ &\mathcal{B}_{m;j_1,0}^{\frac{1}{3},0}\\
a&=3,2,1 &\Delta&=-\frac{1}{3}m, \quad j_1=j_2=0  &\ &\mathcal{B}_{m;0,0}^{\frac{1}{2},0}
\end{align}
according to the number of charges obeying the condition \eqref{shortening}. Obviously we can also consider the conjugate shortening condition 
\begin{equation}
\bar Q_a\ket{\Delta, m, j_1, j_2}^{\text{hw}}=0
\end{equation}
which yields short  multiplets conjugate to the ones considered above
\begin{align}\label{shortening2}
a&=1\qquad &\Delta&=\frac{1}{3}(j_2+2j_1+m)\qquad &\ &\mathcal{B}_{m;j_1,j_2}^{0,\frac{1}{6}}\\
a&=1,2   &\Delta&=\frac{1}{3}(j_2+m),\quad j_1=0 &\ &\mathcal{B}_{m;0,j_2}^{0,\frac{1}{3}}\\
a&=1,2,3 &\Delta&=\frac{1}{3}m, \quad j_1=j_2=0  &\ &\mathcal{B}_{m;0,0}^{0,\frac{1}{2}}
\end{align} 
Finally we may have mixed multiplets where the highest weight is annihilated both by $Q^a$ and $\bar Q_a$. Those include
\begin{align}
 &{\mathcal{B}}^{\frac16 ,\frac16}_{m;j_1,j_2} & \Delta&=\frac{j_2+j_1}{2} & m&=\frac{j_2-j_1}{2}\\
  &{\mathcal{B}}^{\frac13 ,\frac16}_{m;j_1,0} & \Delta&=\frac{j_1}{2} & m&=-\frac{j_1}{2} & &j_2=0\\
  &{\mathcal{B}}^{\frac16, \frac13}_{m;0,j_2} & \Delta&=\frac{j_2}{2} & m&=\frac{j_2}{2} & &j_1=0\label{shortening3}
\end{align}

 \section{Supersymmetry transformations}

\subsection{In $SU(4)$ notations}

The ABJ(M) action in \eqref{action} is invariant under the following superconformal transformations 
\bea \label{susytransf}
\delta C_K&=&- \bar\zeta^{IJ,\alpha}\ \varepsilon_{IJKL}\ \bar \psi^{L}_\alpha \non  \\
\delta\bar C^K&=&2\bar\zeta^{KL,\alpha}\ \psi_{L,\alpha} \non \\
\delta\bar\psi^{K,\beta}&=& 2i\bar\zeta^{KL,\alpha}{(\gamma^\mu)_\alpha}^\beta D_\mu C_L + \frac{4\pi i}{k}\bar\zeta^{KL,\beta}(C_L\bar C^M C_M-C_M\bar C^MC_L) 
+ \frac{8\pi i}{k}\bar\zeta^{IJ,\beta}C_I\bar C^KC_J \non \\
&&\quad + \, 2i\bar\epsilon^{KL,\beta} C_L \notag \non \\
\delta\psi^\beta_K&=&-i\bar\zeta^{IJ,\alpha}\varepsilon_{IJKL}{(\gamma^\mu)_\alpha}^\beta D_\mu \bar C^L+\frac{2\pi i}{k}\bar\zeta^{IJ,\beta}\varepsilon_{IJKL}(\bar C^LC_M\bar C^M-\bar C^MC_M\bar C^L)\non \\
&&\quad +\frac{4\pi  i}{k}\bar\zeta^{IJ,\beta}\varepsilon_{IJML}\bar C^M C_K\bar C^L-i\bar\epsilon^{IJ,\beta}\varepsilon_{IJKL}\bar C^L\non \\
\delta A_\mu&=&\frac{4\pi i}{k}\bar\zeta^{IJ, \alpha}{(\gamma_\mu)_\alpha}^\beta\bigg(C_I\psi_{J\beta} - \frac{1}{2}\varepsilon_{IJKL}\bar\psi^K_\beta\bar C^L\bigg) \non \\
\delta\hat A_\mu&=&\frac{4\pi i}{k}\bar\zeta^{IJ, \alpha}{(\gamma_\mu)_\alpha}^\beta\bigg(\psi_{J\beta}C_I -\frac{1}{2}\varepsilon_{IJKL}\bar C^L\bar\psi^K_\beta\bigg)
\eea
where the parameters of the transformations are expressed in terms of supersymmetry and superconformal parameters as 
\beq
\bar\zeta^{IJ}_{\alpha} = \bar\Theta^{IJ}_{\alpha} - x^\mu (\gamma_\mu)_\a^{\; \b} {\bar \e}^{IJ}_\b
\eeq
We recall that they satisfy $\bar\zeta^{IJ} = - \bar\zeta^{JI}$, and are subject to the reality conditions $\bar\zeta^{IJ} = (\zeta_{IJ})^\ast$ with $\zeta_{IJ} =  \frac12 \e_{IJKL}
\bar\zeta^{KL}$.

If we set ${\bar \e}^{IJ} = 0$ in \eqref{susytransf} we obtain ${\cal N}=6$ supersymmetry transformations. Expressing them as
\beq\label{susytransf2}
\delta \Phi =  [ \bar\Theta^{IJ} \bar{Q}_{IJ}, \Phi ]   =  [\Theta_{IJ} Q^{IJ} , \Phi ] 
\eeq
for a generic field $\Phi$, it is easy to realize that the $Q^{IJ}$ supercharges (or equivalently $\bar{Q}_{IJ}$) satisfy the $\mathfrak{osp}(6|4)$ algebra \eqref{QSalgebra} under the identification $P_\mu = i \partial_\mu$. 

\vskip 10pt

\subsection{In $SU(3)$ notations}\label{sect: su(3)susy}

The generic supersymmetry transformation defined in \eqref{susytransf2} can be specialized to the $\mathfrak{su}(1,1|3)$ supercharges $(Q^a, \bQ_a)$ defined in (\ref{supercharges},\ref{anticomm}). For a generic field $\tilde{\Phi}$ in a given representation of the $\mathfrak{su}(3)$ R-symmetry algebra it reduces to 
\beq\label{susytransf3}
\delta \tilde\Phi =\commutator*{\theta_aQ^a + \btheta^a\bQ_a}{\tilde\Phi} 
\eeq
under the parameter identification 
\bea
&& \theta_a = 2\, \Theta^1_{1(a+1)}  \qquad a=1,2,3  \\
&& \bar{\theta}^{1} = -2i \, \Theta_{34}^2   \qquad \bar{\theta}^{2}  = -2i \, \Theta_{42}^2  \qquad  \bar{\theta}^{3} = -2i \, \Theta_{23}^2  \non 
\eea
From the variations in \eqref{susytransf} we can easily read the supersymmetry transformations of the ABJ(M) fundamental fields reorganized in $\mathfrak{su}(3)$ R-symmetry representations (see  eqs. \eqref{su3breaking} and \eqref{gauge}). Comparing these transformations with the general variation defined in \eqref{susytransf3} we obtain the action of the supercharges on the fields, which takes the following form 
\begin{itemize}
\item Scalar fields
\begin{align} 
Q^aZ&=-\bar{\chi}_1^a 		& \bar{Q}_aZ&=0 		&Q^a\bar{Z}&=0			& \bar{Q}_a\bar{Z}&=i\chi_a^1 \non \\
Q^a Y_b&= \delta^a_b\bar{\psi}_1		&\bar{Q}_aY_b&=- i\epsilon_{abc}\bar{\chi}_2^c		&Q^a\bar{Y}^b&=-\epsilon^{abc}\chi^2_c		&	\bar{Q}_a\bar{Y}^b&=-i\delta_a^b\psi^1
\end{align}

\item Fermions
\begin{subequations}
\begin{align} 
\bar{Q}_a\psi^1&=0			&Q^a\psi^1&= - iD_3\bar{Y}^a - \frac{2\pi i}{k}\left( \bar{Y}^al_B-\hat{l}_B\bar{Y}^a \right) 	\\
Q^a\psi^2&=-i D \bar{Y}^a			&\bar{Q}_a\psi^2&=-\frac{4\pi }{k}\epsilon_{abc}\bar{Y}^bZ\bar{Y}^c	\\
\bar{Q}_a\chi_b^1&=	 \epsilon_{abc} \, \bar{D} \bar{Y}^c		&Q^a\chi_b^1&= i\delta_b^aD_3\bar{Z} + \frac{4\pi i}{k}\left(\bar{Z}\Lambda^a_b-\hat{\Lambda}^a_b\bar{Z}\right)	
\\
Q^a\chi_b^2&=i\delta^a_b \, D \bar{Z}			&\bar{Q}_a\chi_b^2&= -\epsilon_{abc}D_3\bar{Y}^c - \frac{2\pi }{k}\epsilon_{acd}\left( \bar{Y}^c\Theta_b^d-\hat{\Theta}_b^d\bar{Y}^c \right) 	\\
Q^a\bar{\psi}_1&=0			&\bar{Q}_a\bar{\psi}_1&=- D_3 Y_a - \frac{2\pi }{k}\left( Y_a\hat{l}_B-l_B Y_a \right)	\\
\bar{Q}_a\bar{\psi}_2&=- \bar{D}Y_a			&Q^a\bar{\psi}_2&= \frac{4\pi i}{k}\epsilon^{abc}Y_b\bar{Z}Y_c	\\
Q^a\bar{\chi}^b_1&=	- i\epsilon^{abc} \, D Y_c		&\bar{Q}_a\bar{\chi}^b_1&=\delta_b^aD_3Z + \frac{4\pi}{k}\left(Z\hat{\Lambda}^a_b-\Lambda^a_bZ\right)	\\
\bar{Q}_a\bar{\chi}^b_2&= \delta_a^b \, \bar{D}Z			&Q^a\bar{\chi}^b_2&=i\epsilon^{abc}D_3Y_c + \frac{2\pi i}{k}\epsilon^{acd}\left(Y_c\hat{\Theta}_d^b-\Theta_d^b Y_c \right)
\end{align}
\end{subequations}

\item Gauge fields
\begin{equation}
\begin{alignedat}{3} 
Q^aA_3&=- \frac{2\pi i}{k}\left(\bar{\psi}_1\bar{Y}^a-\bar{\chi}_1^a\bZ+\epsilon^{abc}Y_b\chi_c^2 \right)			
&& \quad \bar{Q}_aA_3&&=\frac{2\pi }{k}  \left(Z\chi^1_a-Y_a\psi^1 -\epsilon_{abc}\bar{\chi}_2^b\bar{Y}^c \right) \\
Q^aA&=0		&& \quad	\bar{Q}_a A &&=-\frac{4\pi}{k}\left(Y_a\psi^2-Z\chi_a^2 - \epsilon_{abc}\bar{\chi}_1^b\bar{Y}^c\right)	\\
Q^a \bar{A}&=- \frac{4\pi i}{k}	\left( \bar{\psi}_2\bar{Y}^a-\bar{\chi}^a_2\bZ -\epsilon^{abc}Y_b\chi_c^1\right)		
&&\quad \bar{Q}_a \bar{A} &&=0 \\
Q^a\hat{A}_3&=-\frac{2\pi i}{k}\left(\bar{Y}^a\bar{\psi}_1-\bZ\bar{\chi}_1^a + \epsilon^{abc}\chi_c^2Y_b \right) && \quad \bar{Q}_a\hat{A}_3&&=\frac{2\pi }{k} \left(\chi^1_aZ-\psi^1Y_a - \epsilon_{abc}\bar{Y}^c\bar{\chi}_2^b \right) \\
Q^a \hat{A} &=0		
&& \quad \bar{Q}_a \hat{A} &&=\frac{4\pi}{k}\left(\psi^2Y_a-\chi_a^2Z - \epsilon_{abc}\bar{Y}^c\bar{\chi}_1^b\right)	\\
Q^a \hat{\bar A} &= - \frac{4\pi i}{k}	\left( \bar{Y}^a\bar{\psi}_2-\bZ\bar{\chi}^a_2 - \epsilon^{abc}\chi_c^1Y_b\right) && \quad	
\bar{Q}_a \hat{\bar A} &&=0
\end{alignedat}
\end{equation}
\end{itemize}
where we have defined the bilinear scalar fields
\begin{align}
\begin{pmatrix}
\Lambda_a^b & 0\\
0 & \hat{\Lambda}_a^b
\end{pmatrix}
&=
\begin{pmatrix}
Y_a\bar{Y}^b+\frac{1}{2}\delta_a^bl_B & 0\\
0 & \bar{Y}^b Y_a +\frac{1}{2}\delta_a^b\hat{l}_B 
\end{pmatrix} \non \\
\begin{pmatrix}
\Theta_a^b & 0\\
0 & \hat{\Theta}_a^b
\end{pmatrix}
&=
\begin{pmatrix}
Y_a\bar{Y}^b-\delta_a^b (Z\bZ+Y_c\bY^c) & 0\\
0 & \bar{Y}^b Y_a -\delta_a^b(\bZ Z+\bY^cY_c)
\end{pmatrix} \non \\
\begin{pmatrix}
l_B & 0\\
0 & \hat{l}_B
\end{pmatrix}
&=
\begin{pmatrix}
Z\bZ-Y_c\bY^c & 0\\
0 & \bZ Z-\bY^cY_c
\end{pmatrix}
\end{align}

\vskip25pt

\section{Two-loop integrals}\label{app:twoloop}

 In this appendix we list the integrals corresponding to the two-loop diagrams in figures \ref{scalarcorr}-\ref{threeeyes}, dressed by their color factors. 

Diagram \ref{scalarcorr} contains the two-loop correction to the scalar propagator. This has been computed in \cite{Bianchi:2018bke} and reads
\begin{align}\label{eq:2loopscalar}
 {\cal C}(N_1,N_2) \, \equiv \, & \raisebox{-2.5mm}{\includegraphics[scale=0.3]{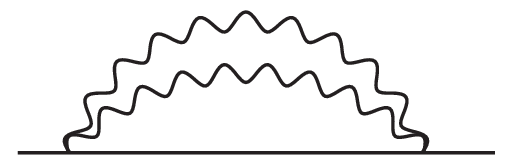}}\, +\, \raisebox{-3.5mm}{\includegraphics[scale=0.3]{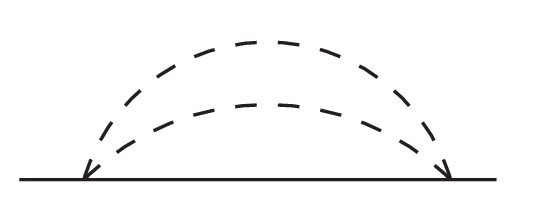}}\, +\, \raisebox{-4mm}{\includegraphics[scale=0.3]{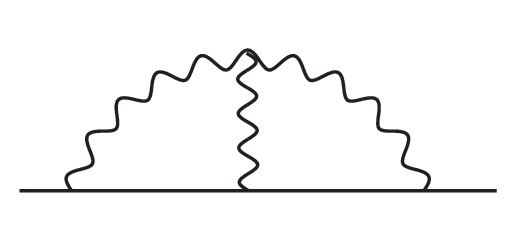}} \nonumber\\&
+ \raisebox{-6.5mm}{\includegraphics[scale=0.3]{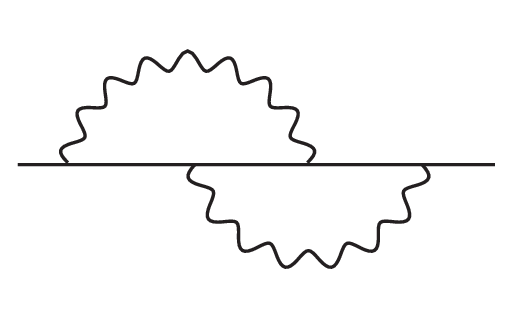}}\, +\, \raisebox{-5mm}{\includegraphics[scale=0.3]{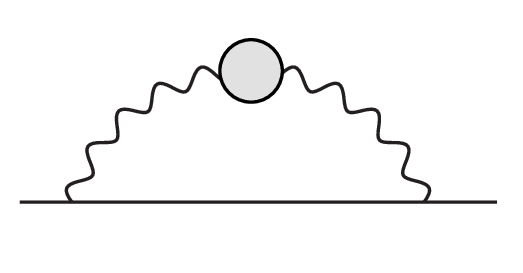}}  \nonumber \\
= & \, \frac{N_1 N_2}{k^2} \,  \left(N_1^2+N_2^2-4 N_1 N_2+2\right)\left(\frac{\pi}{3 \epsilon }+2 \pi +O\left(\epsilon \right)\right) \nonumber \\
& + \frac{N_1 N_2}{k^2} \, \left(N_1^2+N_2^2-2\right)\left(-\frac{4\pi}{3 \epsilon }+\pi  \left(\pi ^2-8\right) +O\left(\epsilon \right)\right) \nonumber \\
& +  \frac{N_1 N_2}{k^2} \, \left(N_1 N_2-1\right)\left(-\frac{8 \pi}{3 \epsilon } + 4\pi(\pi^2 -20 \pi) +O\left(\epsilon \right)\right)
\end{align}

To compute the contributions of the other diagrams it is sufficient to rely on Feynman rules listed in appendix \ref{ABJ(M)}, together with the product of polarization vectors. Explicitly, we find
\begin{equation}
\begin{aligned}
\eqref{kite} & = - s^2 \; \frac{\Gamma^6\left(\frac{1}{2}-\epsilon\right)}{32 \, \pi^{7-6\epsilon}} \, \frac{N_1^2 N_2^2}{k^2}  \int d^dx d^dy \frac{x^\mu y^\nu}{(x^2)^{\frac{3}{2}-\epsilon}(y^2)^{\frac{3}{2}-\epsilon}\left((x-s)^2\right)^{\frac{1}{2}-\epsilon}\left((y-s)^2\right)^{\frac{1}{2}-\epsilon}}  \\
& \hspace{5cm} \times \; \left[\frac{\delta_{\mu\nu}}{\left[\left(x-y\right)^2\right]^{1-2\epsilon}}-\de_\mu\de_\nu\frac{\left[(x-y)^2\right]^{2\epsilon}}{4\epsilon(1+2\epsilon)}\right]
\end{aligned}
\label{}
\end{equation}

\vskip 10pt
\begin{equation}
\begin{aligned}
&\eqref{doublegauge} =s^2 \; \frac{\Gamma^4\left(\frac{1}{2}-\epsilon\right)\Gamma^2\left(\frac{3}{2}-\epsilon\right)}{128 \, \pi^{7-6\epsilon}}  \frac{N_1 N_2}{k^2} ((N_1 - N_2)^2 - 2 N_1 N_2 +2) \; \times\\
	&\qquad \quad \int\frac{d^dx d^dy}{(x^2)^{\frac{1}{2}-\epsilon}(y^2)^{\frac{1}{2}-\epsilon}\left((x-y)^2\right)^{2-2\epsilon}\left((x-s)^2\right)^{\frac{1}{2}-\epsilon}\left((y-s)^2\right)^{\frac{1}{2}-\epsilon}}
\end{aligned}
\label{label}
\end{equation}

\vskip 10pt
\begin{equation}
\begin{aligned}
&\eqref{duobleph}=   s^2 \; \frac{\Gamma^6\left(\frac{1}{2}-\epsilon\right)\Gamma^2\left(\frac{3}{2}-\epsilon\right)}{256 \, \pi^{10-8\epsilon}} \;
\frac{N_1 N_2}{k^2} \, \left( N_1 - N_2\right)^2  \;   \varepsilon_{\mu\nu\eta} \varepsilon_{\rho \sigma  \tau} \; \times \\
& \hspace{-0.5cm} \int d^dx d^dy d^dz d^dw \, \frac{(x-y)^\eta (z-w)^\tau}{\left((x-y)^2\right)^{\frac{3}{2}-\epsilon}\left((z-w)^2\right)^{\frac{3}{2}-\epsilon}\left((x-s)^2\right)^{\frac12-\epsilon} \left((y-s)^2\right)^{\frac12-\epsilon} (z^2)^{\frac12-\epsilon} (w^2)^{\frac12-\epsilon}} \\
& \qquad \qquad \qquad \times \; \partial^\mu \partial^\rho \frac{1}{\left( (x-z)^2\right)^{\frac12 - \epsilon}} \, \partial^\nu \partial^\sigma \frac{1}{\left((y-w)^2\right)^{\frac12 - \epsilon}}
\end{aligned}
\label{}
\end{equation}

\vskip 10pt
\begin{equation}
\begin{aligned}
&\eqref{crossedduobleph} =  - s^2 \; \frac{\Gamma^6\left(\frac{1}{2}-\epsilon\right)\Gamma^2\left(\frac{3}{2}-\epsilon\right)}{128 \, \pi^{10-8\epsilon}} \;
\frac{N_1 N_2}{k^2} \, \left( N_1N_2 -1 \right) \;  \varepsilon_{\mu\nu\eta} \varepsilon_{\rho \sigma  \tau} \; \times \\
& \hspace{-0.5cm}  \int d^dx d^dy d^dz d^dw \, \frac{(x-y)^\eta (z-w)^\tau}{\left((x-y)^2\right)^{\frac{3}{2}-\epsilon}\left((z-w)^2\right)^{\frac{3}{2}-\epsilon} \left((x-s)^2\right)^{\frac12-\epsilon} \left((w-s)^2\right)^{\frac12-\epsilon} (y^2)^{\frac12-\epsilon} (z^2)^{\frac12-\epsilon}} \\ 
&  \qquad \qquad \qquad \times \;\partial^\mu \partial^\rho \frac{1}{\left( (x-z)^2\right)^{\frac12 - \epsilon}} \, \partial^\nu \partial^\sigma \frac{1}{\left((y-w)^2\right)^{\frac12 - \epsilon}} 
\end{aligned}
\label{}
\end{equation}

\vskip 10pt
\begin{equation}
\begin{aligned}
& \eqref{4ptccff} =  s^2 \; \frac{\Gamma^4\left(\frac{1}{2}-\epsilon\right)\Gamma^2\left(\frac{3}{2}-\epsilon\right)}{16 \, \pi^{7-6\epsilon}}  \frac{N_1 N_2}{k^2} (N_1N_2 -1)  \; \times\\
	&\qquad \quad \int\frac{d^dx d^dy}{(x^2)^{\frac{1}{2}-\epsilon}(y^2)^{\frac{1}{2}-\epsilon}\left((x-y)^2\right)^{2-2\epsilon}\left((x-s)^2\right)^{\frac{1}{2}-\epsilon}\left((y-s)^2\right)^{\frac{1}{2}-\epsilon}}
\end{aligned}
\label{}
\end{equation}

We note that in the large $N_1,N_2$ approximation we obtain $ \eqref{4ptccff} = - 4 \eqref{doublegauge}$, in agreement with the results in \cite{Young:2014lka}.

\begin{equation}
\begin{aligned}
&\eqref{triangle}=  -s^2 \;  \frac{\Gamma^5\left(\frac{1}{2}-\epsilon\right)\Gamma^2\left(\frac{3}{2}-\epsilon\right)}{128 \, \pi^{\frac{17}{2}-7\epsilon}} \; \frac{N_1 N_2}{k^2} (N_1^2+ N_2^2 - 4 N_1 N_2 +2)  \; \varepsilon_{\mu \rho \sigma} \varepsilon_{\mu \nu \eta} \; \times \\ 
&  \int d^dx d^dy d^dz \frac{(x-z)^\sigma}{((x-z)^2)^{\frac{3}{2}-\epsilon}} \,  \frac{(x-y)^\eta}{((x-y)^2)^{\frac{3}{2}-\epsilon}} \; \times \\
&\qquad \quad \partial^\rho \frac{1}{((y-z)^2)^{\frac{1}{2}-\epsilon}} \; \partial^\nu \frac{1}{((y-s)^2)^{\frac{1}{2}-\epsilon}} \; \frac{1}{(x^2)^{\frac{1}{2}-\epsilon}(z^2)^{\frac{1}{2}-\epsilon}((x-s)^2)^{\frac{1}{2}-\epsilon}}
\end{aligned}
\label{}
\end{equation}

\vskip 10pt
\begin{equation}
\eqref{fermgauge}=0 \qquad \qquad \qquad \eqref{gaugethree}=0
\label{}
\end{equation}

\vskip 10pt
\begin{equation}
\begin{aligned}
& \eqref{fork}=  s^2 \;  \frac{\Gamma^5\left(\frac{1}{2}-\epsilon\right)\Gamma^3\left(\frac{3}{2}-\epsilon\right)}{128 \, \pi^{10-8\epsilon}} \; \frac{N_1 N_2}{k^2} (N_1^2 + N_2^2 -2)  \; \varepsilon_{\rho \nu \tau}  \varepsilon_{\rho \eta \sigma} \varepsilon_{\nu\mu \varphi} \varepsilon_{\tau \chi\xi} \; \times \\
&  \, \int d^dx  d^dy d^dz d^dw \frac{(x-z)^\varphi (y-z)^\xi (w-z)^\sigma}{\left( (x-z)^2 \right)^{\frac32 - \epsilon} \left( (y-z)^2 \right)^{\frac32 - \epsilon} \left( (w-z)^2 \right)^{\frac32 - \epsilon} } \\
& \qquad \times \;  \partial^\eta \frac{1}{\left( (w-s)^2\right)^{\frac12 -\epsilon}} \; \partial^\chi  \frac{1}{\left( (x-y)^2 \right)^{\frac12 -\epsilon}} \;   \partial^\mu \frac{1}{\left( (x-s)^2 \right)^{\frac12 -\epsilon}} \; \, \frac{1}{ (y^2)^{\frac12 -\epsilon} ( w^2 )^{\frac12 -\epsilon} }
\end{aligned}
\label{}
\end{equation}

\vskip 10pt
\begin{equation}
\begin{aligned}
& \eqref{scalarfork}=  s^2 \; \frac{\Gamma^6\left(\frac{1}{2}-\epsilon\right)\Gamma^2\left(\frac{3}{2}-\epsilon\right)}{256 \, \pi^{10-8\epsilon}} \; 
\frac{N_1 N_2}{k^2} (N_1N_2-2) \;  \varepsilon_{\mu \nu \epsilon}  \varepsilon_{\rho\sigma\tau} \; \times \\
& \qquad    \int d^dx  d^dy d^dz d^dw \frac{(x-y)^\epsilon (z-w)^\tau }{\left( (x-y)^2 \right)^{\frac32 - \epsilon} \left( (z-w)^2 \right)^{\frac32 - \epsilon}} \,  \frac{1}{\left( (w-s)^2 \right)^{\frac12 -\epsilon}}  \\
& \qquad \; \times  \; \partial^\rho \frac{1}{\left( (x-z)^2 \right)^{\frac12 -\epsilon}} \, \partial^\nu \frac{1}{\left( (y-z)^2 \right)^{\frac12 -\epsilon}} \partial^\sigma \frac{1}{\left( w^2 \right)^{\frac12 -\epsilon}}  \\
& \hspace{3cm}\; \times  \; \left[ \partial^\mu \frac{1}{\left( (x-s)^2 \right)^{\frac12 -\epsilon}} \, \frac{1}{\left( y^2 \right)^{\frac12 -\epsilon}}  - 
\partial^\mu \frac{1}{\left( x^2 \right)^{\frac12 -\epsilon}} \, \frac{1}{\left( (y-s)^2 \right)^{\frac12 -\epsilon}} \right]
\end{aligned}
\end{equation}

\vskip 10pt
\begin{equation}
\begin{aligned}
\eqref{threeeyes} & = - s^2 \; \frac{\Gamma^4\left(\frac{1}{2}-\epsilon\right)\Gamma^2\left(\frac{3}{2}-\epsilon\right)}{32 \, \pi^{7-6\epsilon}}  \frac{N_1 N_2}{k^2} (N_1- N_2)^2  \\ 
& \qquad \qquad \qquad  \times \;  \int\frac{d^dx d^dy}{\left((x-s)^2\right)^{1-2\epsilon} \left((x-y)^2\right)^{2-2\epsilon}\left(y^2\right)^{1-2\epsilon}}
\end{aligned}
\end{equation}

\newpage

\bibliography{biblio}

\end{document}